\documentclass[11pt]{article}
\usepackage{bbm}
\usepackage{stmaryrd}
\usepackage{amsmath}
\usepackage{amsfonts}
\usepackage{latexsym,amssymb,amsmath,mathrsfs,amsbsy, amsthm}
\usepackage[usenames]{color}
\usepackage{xspace,colortbl}

\newtheorem{thm}{Theorem}[section]
\newtheorem{lem}[thm]{Lemma}

\newtheorem{prop}[thm]{Proposition}

\newcommand{\md}{\mathrm{d}}
\newcommand{\x}{\mathcal {X}}
\newcommand{\y}{\mathcal {Y}}
\newcommand{\z}{\mathcal {Z}}
\newcommand{\U}{\mathcal {U}}
\newcommand{\V}{\mathcal {V}}
\newcommand{\W}{\mathcal {W}}
\newcommand{\h}{\mathcal {H}}
\newcommand{\psp}{\vspace{0.4cm}}
\numberwithin{equation}{section}

\begin{document}

\title{\textbf{Asymmetric and Moving-Frame \\ Approaches  to MHD Equations} }

\author{Bintao Cao\footnote{Institute of Mathematics, Academy of Mathematics and Systems Science,
Chinese Academy of Sciences, Beijing 100190, P. R. China,
  caobintao@amss.ac.cn} \ }
\date{}
\maketitle

\begin{abstract}
    The magnetohydrodynamic (MHD) equations of incompressible viscous
fluids with finite electrical conductivity describe the motion of
viscous electrically conducting fluids in a magnetic field. In this
paper, we find twelve families of solutions of these equations by
Xu's asymmetric and moving frame methods. A family of singular
solutions may reflect basic characteristics of vortices. The other
solutions are globally analytic with respect to the spacial
variables. In particular,   Bernoulli equation and Wronskian
determinants play important roles in our approaches. Our solutions
may also help engineers to develop more effective algorithms to find
physical numeric solutions to practical models.
\end{abstract}


\vskip 0.2cm


\renewcommand{\theequation}{\thesection.\arabic{equation}}
\setcounter{equation}{0}

\section{Introduction}

In physics, magnetohydrodynamic (MHD) studies the motion of
electrically conducting fluids in a magnetic field. Examples are
astrophysical plasma, solar atmosphere, magnetosphere, thermonuclear
fusion and liquid metal etc. Their motion described by the MHD
equations (see \cite{GLL} for the details), which coupled the
Navier-Stokes equations and the Maxwell's equations.

There are many different approaches to the MHD equations. The
wellposedness of these equations has been well studied (e.g., cf.
\cite{CCM,CMPY,CMZ,W1,W2,Z}). Numerical methods for compressible
case were developed in \cite{BW,ZC,RB,WD,CKC,L,M,T,PRLGZ,LD,R}, etc.
One can also find similar results for the incompressible case in
\cite{GLL,GMP,HSS,S,P,KK}. The Lagrange invariance of a generalized
ion magnetic helicity was established for Hall MHD in \cite{Sh}.
Vod\'{a}k \cite{V} proved the existence of a weak, local in time
solution to the compressible case on the whole space under special
assumptions on pressure and entropy. For the historic introduction
of the MHD equations, we refer the Section 1 of \cite{C}.

In this paper, we investigate an incompressible case, that is, the
MHD equations of incompressible viscous fluids with finite
electrical conductivity (e.g., cf. \cite{Ib}):
\begin{equation}\label{M1}
\nabla\cdot\textbf{v}=0,
\end{equation}
\begin{equation}\label{M2}
\textbf{v}_t+(\textbf{v}\cdot\nabla)\textbf{v}-\mu_0(\textbf{H}\times\textrm{rot}\textbf{H})+\frac{1}{\rho}\nabla
p=\nu \Delta\textbf{v},
\end{equation}
\begin{equation}\label{M3}
\nabla\cdot\textbf{H}=0,
\end{equation}
\begin{equation}\label{M4}
\textbf{H}_t=\textrm{rot}(\textbf{v}\times\textbf{H})+\eta\Delta\textbf{H}.
\end{equation}
Here $\textbf{v}=(u,v,w)^t$ is the velocity vector,
$\textbf{H}=(H^1,H^2,H^3)^t$ is the magnetic field vector, $\nu$ is
the viscosity coefficient, $\mu_0$ is the magnetic permeability and
$\eta$ is the magnetic viscosity. The superscript ``$t$'' denotes
the transpose of a matrix or a vector.

Nucci \cite{Nu} found the Lie point symmetries of these equations.
Popovych \cite{Po1,Po2,Po3} reduced \eqref{M1}-\eqref{M4} to
ordinary differential equations and to partial differential
equations in two and in three independent variables by means of Lie
symmetries. Then he constructed some classes of similarity
solutions.

In order to solve the three-dimensional Navier-Stokes equations, Xu
\cite{X1} introduced a method of imposing asymmetric conditions on
the velocity vector with respect to independent variables and a
method of moving frame which reflect the rotation nature of flow. By
these methods, he found seven families of non-steady rotating
asymmetric solutions with multiple parameter functions. In
particular, a family of singular solutions may reflect important
features of vortices. The other solutions are globally analytic with
respect to spacial variables. The goal of this work is to find
similar solutions for the MHD equations  \eqref{M1}-\eqref{M4}.

Note that when ${\bf H}=\vec 0$, the MHD equations
\eqref{M1}-\eqref{M4} become the three-dimensional Navier-Stokes
equations. The factor of magnetic field makes the MHD equations
much more complicated than the Navier-Stokes equations. For
instance, Xu \cite{X1} reduced the Navier-Stokes equations under
certain conditions to a linear differential equation, and we can
only reduce the MHD equations to a system of linear differential
equations. In particular, we have to use Wronskian determinants to
solve such systems. Moreover, the magnetic field also naturally
leads the appearance of Bernoulli equation in our reductions.

Since the system of  MHD equations  is a coupled  system of
Navier-Stokes equations and the Maxwell's equations, Xu's methods
and exact solutions of the Navier-Stokes equations do give us hints
of how to solve the MHD equations exactly. We find twelve families
of solutions of the equations with multiple parameter functions. One
of them is  a family of singular solutions that may reflect
important characteristics of vortices. The other solutions are
globally analytic with respect to the spacial variables, related to
polynomials, exponential functions or trigonometric functions. Our
solutions may help mathematicians to develop the method of proving
the well known regularity of the equations. They may also help
engineers to develop more effective algorithms to find physical
numeric solutions to practical models. Below we give more technical
details.

For convenience, we always assume that all the involved partial
derivatives of related functions always exist and we can change
the order of taking partial derivatives. We also use prime $'$ to
denote the derivative of any one-variable function.

According to \cite{Nu}, the  MHD equations are invariant under the
following transformations:

1. Time translation:
\begin{equation}\label{T1}
T_1:\ \ \ \ t\mapsto t+a,
\end{equation} where $a$ is an arbitrary real constant.

2. Scale transformation:
\begin{equation}\label{T2}
T_2:\ \ \ \ (t,\textbf{x},\textbf{v},\textbf{H},p)\mapsto
(\lambda^2t,\lambda\textbf{x},\lambda\textbf{v},\lambda^{-1}\textbf{H},\lambda^{-2}p),
\end{equation} where $\lambda$ is any nonzero real constant and  $\textbf{x}=(x,y,z)^t$.

3. Rotations:
\begin{equation}\label{T3}
T_3:\ \ \ \ (t,\textbf{x},\textbf{v},\textbf{H},p)\mapsto
(t,R\textbf{x},R\textbf{v},R\textbf{H},p)
\end{equation} for any $R\in SO(3)$.

4. Moving coordinates:
\begin{equation}\label{T4}
T_4:\ \ \ \ (t,\textbf{x},\textbf{v},\textbf{H},p)\mapsto
(t,\textbf{x}+\textbf{F}(t),\textbf{v}+\textbf{F}'(t),\textbf{H},
p-\rho\textbf{x}\cdot\textbf{F}''),
\end{equation} where $\textbf{F}(t)=(f(t),g(t),h(t))^t$ is an
arbitrary vector valued functions.

5. Pressure changes:
\begin{equation}\label{T5}
T_5:\ \ \ \ (t,\textbf{x},\textbf{v},\textbf{H},p)\mapsto
(t,\textbf{x},\textbf{v},\textbf{H},p+\theta(t))
\end{equation} for any function $\theta$ of $t$.

The above transformations transform a solution of the  MHD equations
into other solutions. So we only need to find exact solutions modulo
these transformations.

Assuming
\begin{equation}\label{E1}
   \left\{\begin{aligned}
     u&=\frac{\alpha}{2}x+y\phi(t,x^2+y^2),\\
     v&=\frac{\alpha}{2}y-x\phi(t,x^2+y^2),\\
     w&=-\alpha z+\psi(t,x^2+y^2),
   \end{aligned}
   \right.
\ \ \ \ \ \ \ \ \ \ \ \
   \left\{\begin{aligned}
     H^1&=\frac{\beta}{2}x+y\xi(t,x^2+y^2),\\
     H^2&=\frac{\beta}{2}y-x\xi(t,x^2+y^2),\\
     H^3&=-\beta z+\sigma(t,x^2+y^2)
   \end{aligned}
   \right.
\end{equation} for some functions $\alpha$ and $\beta$ of $t$, and
two-variable functions $\phi,\psi,\xi,\sigma$,
  we obtain the following solution:
\begin{equation}
   \left\{\begin{aligned}
     u&=\frac{\alpha}{2}x+y\sum_{m=0}^n a_m(t)(x^2+y^2)^m,\\
     v&=\frac{\alpha}{2}y-x\sum_{m=0}^n a_m(t)(x^2+y^2)^m,\\
     w&=-\alpha z+\sum_{m=0}^n g_m(t)(x^2+y^2)^m,
   \end{aligned}
   \right.
\ \ \ \ \ \ \ \ \ \ \ \
   \left\{\begin{aligned}
     H^1&=\frac{b}{2}x+y\sum_{m=0}^nc_m(t)(x^2+y^2)^m,\\
     H^2&=\frac{b}{2}y-x\sum_{m=0}^nc_m(t)(x^2+y^2)^m,\\
     H^3&=-b z+\sum_{m=0}^nf_m(t)(x^2+y^2)^m.
   \end{aligned}
   \right.
\end{equation} and $p$ is given by \eqref{3.86}, where $n$ is any
given positive integer,
\begin{equation} \alpha=\frac{k+2k^2le^{-kt}}{-1+2kle^{-kt}}
\end{equation} with constants $k$ and $l$ is obtained by solving
a related Bernoulli equation, $b$ is an arbitrary real constant,
 and $a_m(t), c_m(t), f_m(t), g_m(t)$ are determined via various
 Wronskian determinants (cf. \eqref{3.76}-\eqref{3.85}). This is an example of
 asymmetric approach.

 By Xu's moving-frame approach,  we obtain the following solution of the MHD equations \eqref{M1}-\eqref{M4}:
  \begin{equation}
  \left\{
  \begin{aligned}
    u=&-(\frac{\alpha''}{2\alpha'}+\frac{\mu_0
         ab}{4\alpha'})x-\alpha'y-\frac{6\nu\cos\alpha}{x\cos\alpha+y\sin\alpha}
         +6\nu\sin\alpha\frac{-x\sin\alpha+y\cos\alpha}{(x\cos\alpha+y\sin\alpha)^2},\\
    v=&\alpha'x-(\frac{\alpha''}{2\alpha'}+\frac{\mu_0
         ab}{4\alpha'})y-\frac{6\nu\sin\alpha}{x\cos\alpha+y\sin\alpha}-
         6\nu\cos\alpha\frac{-x\sin\alpha+y\cos\alpha}{(x\cos\alpha+y\sin\alpha)^2},\\
    w=&2(\frac{\alpha''}{2\alpha'}+\frac{\mu_0
         ab}{4\alpha'})z,
  \end{aligned}\right.
  \end{equation}
  \begin{equation} \left\{
  \begin{aligned}
  H^1&=-[(a\cos\alpha-b\sin\alpha)x+(-a\sin\alpha+b\cos\alpha)y]\sin\alpha,\\
  H^2&=[(a\cos\alpha-b\sin\alpha)x+(a\sin\alpha+b\cos\alpha)y]\cos\alpha,\\
  H^3&=-bz
  \end{aligned}\right.\end{equation} and
  \begin{eqnarray}\label{4.61}
  p&=&-\rho\Big(((\frac{\alpha''}{2\alpha'}+\frac{\mu_0
         ab}{4\alpha'})'+(\frac{\alpha''}{2\alpha'}+\frac{\mu_0
         ab}{4\alpha'})^2-(\alpha')^2-\mu_0a^2)\frac{(x\cos\alpha+y\sin\alpha)^2}{2}\nonumber\\
         &&+((\frac{\alpha''}{2\alpha'}+\frac{\mu_0
         ab}{4\alpha'})'+(\frac{\alpha''}{2\alpha'}+\frac{\mu_0
         ab}{4\alpha'})^2-(\alpha')^2)\frac{(-x\sin\alpha+y\cos\alpha)^2}{2}\nonumber\\
      &&-((\frac{\alpha''}{2\alpha'}+\frac{\mu_0
         ab}{4\alpha'})'+2(\frac{\alpha''}{2\alpha'}+\frac{\mu_0
         ab}{4\alpha'})^2)\frac{z^2}{2}\nonumber\\
      &&+(\alpha''+2\alpha'(\frac{\alpha''}{2\alpha'}+\frac{\mu_0
         ab}{4\alpha'}))(x\cos\alpha+y\sin\alpha)(-x\sin\alpha+y\cos\alpha)\nonumber\\
         &&-12\nu\alpha'\frac{-x\sin\alpha+y\cos\alpha}
         {x\cos\alpha+y\sin\alpha}+\frac{12\nu^2}
         {(x\cos\alpha+y\sin\alpha)^2}\Big),
  \end{eqnarray}
where $\alpha$ is any function of $t$, and $a,b$ are arbitrary real
constants. The above solution may partially reflect important
features of vortices.

 The paper is organized as follows. In order to get initial
 feeling of the MHD equations, we solve the equations in Section 2 by assuming
 that the solutions are linear in the
spacial variables. Asymmetric approaches are presented in Section
3. In Section 4, we obtain five families solutions by the
moving-frame method.

\section{Linear solutions}

In this section, we find four families of solutions of the MHD
equations \eqref{M1}-\eqref{M4}, which are linear with respect to
the spacial variables $x$, $y$ and $z$.

Denote
\begin{equation}\label{2.1}
W_1=u_t+uu_x+vu_y+wu_z-\mu_0(H^2(H^2_x-H^1_y)-H^3(H^1_z-H^3_x))-\nu\Delta
u,
\end{equation}
\begin{equation}\label{2.2}
W_2=v_t+uv_x+vv_y+wv_z-\mu_0(H^3(H^3_y-H^2_z)-H^1(H^2_x-H^1_y))-\nu\Delta
v,
\end{equation}
\begin{equation}\label{2.3}
W_3=w_t+uw_x+vw_y+ww_z-\mu_0(H^1(H^1_z-H^3_x)-H^2(H^3_y-H^2_z))-\nu\Delta
w.
\end{equation}
To solve \eqref{M2}, we first deal with the compatibility equations
\begin{equation}\label{2.4}
W_{1y}=W_{2x},\ \ W_{1z}=W_{3x},\ \ W_{2z}=W_{3y}
\end{equation}
and then find $p$. The equations \eqref{M4} can be written as:
\begin{equation}\label{H1}
 H_t^1-(uH^2-vH^1)_y+(wH^1-uH^3)_z-\eta\Delta H_1=0,
\end{equation}
\begin{equation}\label{H2}
 H_t^2-(vH^3-wH^2)_z+(uH^2-vH^1)_x-\eta\Delta H_2=0,
\end{equation}
\begin{equation}\label{H3}
 H_t^3-(wH^1-uH^3)_x+(vH^3-wH^2)_y-\eta\Delta H_3=0.
\end{equation}

\vskip 1.0cm

Let
\begin{equation}\label{2.5}
     \textbf{v}=A\begin{pmatrix}
                  x\\ y\\ z
                  \end{pmatrix},\ \ \ \
     \textbf{H}=B\begin{pmatrix}
                  x\\ y\\ z
                  \end{pmatrix},
\end{equation}
where $A=(a_{i,j}(t))_{3\times3}$ and $B=(b_{i,j}(t))_{3\times3}$
are $3\times 3$ matrices whose entries are functions in $t$. Then by
\eqref{M1} and \eqref{M3}, $\textrm{tr}(A)=\textrm{tr}(B)=0$. Now
the equations \eqref{2.4} become
\begin{equation} \label{2.6}
   \left\{ \begin{aligned}
         &a_{12}'-a_{21}'+(a_{11}+a_{22})(a_{12}-a_{21})+a_{13}a_{32}-a_{23}a_{31}\\
                 &\ \ \ \ +\mu_0((b_{11}+b_{22})(b_{12}-b_{21})+b_{13}b_{32}-b_{23}b_{31}) =0, \\
         &a_{13}'-a_{31}'+(a_{11}+a_{33})(a_{13}-a_{31})+a_{23}a_{12}-a_{21}a_{32}\\
                 &\ \ \ \ +\mu_0((b_{11}+b_{33})(b_{13}-b_{31})+b_{12}b_{23}-b_{21}b_{32})=0, \\
         &a_{23}'-a_{32}'+(a_{22}+a_{33})(a_{23}-a_{32})+a_{13}a_{21}-a_{12}a_{31}\\
                 &\ \ \ \ +\mu_0((b_{22}+b_{33})(b_{23}-b_{32})+b_{13}b_{21}-b_{12}b_{31})
                 =0.
    \end{aligned} \right.
\end{equation}
Moreover, the equations \eqref{M4} are equivalent to the following
system of ordinary differential equations:
\begin{equation}\label{2.7}
   \left\{ \begin{aligned}
                  b_{11}'&=a_{12}b_{21}-a_{21}b_{12}+a_{13}b_{31}-a_{31}b_{13},\\
                  b_{12}'&=a_{11}b_{12}-a_{12}b_{11}+a_{12}b_{22}-a_{22}b_{12}+a_{13}b_{32}-a_{32}b_{13},\\
                  b_{13}'&=2a_{11}b_{13}-2a_{13}b_{11}+a_{22}b_{13}-a_{13}b_{22}+a_{12}b_{23}-a_{23}b_{12},
   \end{aligned} \right.
\end{equation}
\begin{equation}\label{2.8}
   \left\{ \begin{aligned}
                 b_{21}'&=-a_{11}b_{21}+a_{21}b_{11}+a_{22}b_{21}-a_{21}b_{22}+a_{23}b_{31}-a_{31}b_{23},\\
                 b_{22}'&=-a_{12}b_{21}+a_{21}b_{12}+a_{23}b_{32}-a_{32}b_{23},\\
                 b_{23}'&=a_{11}b_{23}-a_{23}b_{11}+2a_{22}b_{23}-2a_{23}b_{22}-a_{13}b_{21}+a_{21}b_{13}
   \end{aligned} \right.
\end{equation}
and
\begin{equation}\label{2.9}
   \left\{ \begin{aligned}
                b_{31}'&=-2a_{11}b_{31}+2a_{31}b_{11}-a_{22}b_{31}+a_{31}b_{22}-a_{21}b_{32}+a_{32}b_{21},\\
                b_{32}'&=a_{11}b_{32}+a_{32}b_{11}-a_{12}b_{31}+a_{31}b_{12}-2a_{22}b_{32}+2a_{32}b_{22},\\
                b_{33}'&=-a_{13}b_{31}+a_{31}b_{13}-a_{23}b_{32}+a_{32}b_{23}.
   \end{aligned} \right.
\end{equation}

We consider the following special cases:\psp

{\it Case 1: $A=A^t$, $B=B^t$.}\psp

Easily see that  the system \eqref{2.6} holds naturally. By
\eqref{2.7}-\eqref{2.9}, we find that $B$ is a constant matrix and
$A$ is determined by following equation:
\begin{equation}\label{2.10}
   \begin{pmatrix}
       b_{12}   & -b_{11}+b_{22} & b_{23}          & -b_{32} & -b_{13}        \\
       2b_{13}  & b_{23}         & -2b_{11}-b_{22} & b_{13}  & -b_{12}        \\
       b_{23}   & b_{13}         & -b_{12}         & b_{23}  & -b_{11}-2b_{22}
   \end{pmatrix}
   \begin{pmatrix}
       a_{11}(t) \\ a_{12}(t) \\ a_{13}(t) \\ a_{22}(t) \\ a_{23}(t)
   \end{pmatrix}=0.
\end{equation}
Note that above system has nontrivial solutions. They span a linear
space. Since
\begin{equation}\label{2.11}
p_x=-\rho W_1,\ \ p_y=-\rho W_2, \ \ p_z=-\rho W_3,
\end{equation}
one gets
\begin{eqnarray}\label{2.12}
     p&=&-\frac{\rho}{2}((a_{11}'+\sum_{j=1}^3a_{1j}^2)x^2+(a_{12}'+\sum_{j=1}^3a_{2j}^2)y^2+(a_{33}'+\sum_{j=1}^3a_{3j}^2)z^2)\nonumber\\
       &&-\rho((a_{12}'+\sum_{j=1}^3a_{1j}a_{2j})xy+(a_{13}'+\sum_{j=1}^3a_{1j}a_{3j})xz+(a_{23}'+\sum_{j=1}^3a_{2j}a_{3j})yz),
\end{eqnarray}
modulo the transformation in \eqref{T5}.

\begin{prop} Let $B$ be any $3\times 3$ constant symmetric matrix
with zero trace and let $A$ be a symmetric matrix function with zero
trace determined by \eqref{2.10}. Then we have the following
solutions of the MHD equations \eqref{M1}-\eqref{M4}:
\begin{equation}\label{2.13}
     \textbf{v}=A\begin{pmatrix}
                  x\\ y\\ z
                  \end{pmatrix},\ \ \ \
     \textbf{H}=B\begin{pmatrix}
                  x\\ y\\ z
                  \end{pmatrix}
\end{equation} and $p$ is given by \eqref{2.12}.
\end{prop}
\psp

 {\it Case 2: $A=A^t$, $B=-B^t$.}\psp

As in case 1, one shows that $B$ is constant and
\begin{equation}\label{2.14}
   a_{11}=\frac{b_{12}^2+b_{13}^2-2b_{23}^2}{3b_{12}b_{13}b_{23}}g(t),\
   \
   a_{22}=\frac{b_{12}^2-2b_{13}^2+b_{23}^2}{3b_{12}b_{13}b_{23}}g(t),
\end{equation}
\begin{equation}\label{2.15}
   a_{12}=\frac{g(t)}{b_{12}},\ \ a_{13}=-\frac{g(t)}{b_{13}},\ \
   a_{23}=\frac{g(t)}{b_{23}},
\end{equation}
where $g$ is an arbitrary function of $t$.

Moreover,
\begin{eqnarray}\label{2.16}
   &&p=-\frac{\rho}{2}((a_{11}'+\sum_{j=1}^3a_{1j}^2+2\mu_0(b_{12}^2+b_{13}^2))x^2+
       (a_{12}'+\sum_{j=1}^3a_{2j}^2+2\mu_0(b_{12}^2+b_{23}^2))y^2\nonumber\\
    &+&(a_{13}'+\sum_{j=1}^3a_{3j}^2+2\mu_0(b_{13}^2+b_{23}^2))z^2)
      -\rho((a_{12}'+\sum_{j=1}^3a_{1j}a_{2j}-2\mu_0b_{13}b_{23})xy\nonumber\\
    &+&(a_{13}'+\sum_{j=1}^3a_{1j}a_{3j}+2\mu_0b_{12}b_{23})xz+
       (a_{23}'+\sum_{j=1}^3a_{2j}a_{3j}-2\mu_0b_{12}b_{13})yz)
\end{eqnarray}
modulo the transformation in \eqref{T5}.

\begin{prop} Let $B$ be a $3\times 3$ constant  skew-symmetric matrix
and let $A$ be a symmetric matrix function with zero trace
determined  by \eqref{2.14} and \eqref{2.15}. Then we have the
following solutions of the MHD equations \eqref{M1}-\eqref{M4}:
\begin{equation}\label{2.17}
     \textbf{v}=A\begin{pmatrix}
                  x\\ y\\ z
                  \end{pmatrix},\ \ \ \
     \textbf{H}=B\begin{pmatrix}
                  x\\ y\\ z
                  \end{pmatrix}
\end{equation} and $p$ is given by \eqref{2.16}.

\end{prop}

{\it Case 3: $A=-A^t$, $B=B^t$.}\psp

One shows that $A$ is constant and
\begin{equation}\label{2.18}
(b_{11},b_{12},b_{13},b_{22},b_{23})^t=\exp(Mt)\textbf{c}
\end{equation}
by \eqref{2.6}-\eqref{2.9}, where $\textbf{c}$ is any
five-dimensional constant vector and
\begin{equation}\label{2.19}
   M=\begin{pmatrix}
       0         &  2a_{12} & 2a_{13} & 0        & 0 \\
       -2a_{12}  &  0       & a_{23}  & 2a_{12}  & a_{13}\\
       -2a_{13}  &  -a_{23} & 0       & -a_{13}  & a_{12}\\
       0         &  -2a_{12}& 0       & 0        & 2a_{23}\\
       -a_{23}    &  -a_{13} & -a_{12} & -2a_{23} & 0
   \end{pmatrix}.
\end{equation}
Moreover,
\begin{eqnarray}\label{2.20}
    p&=&-\rho(-(a_{12}^2+a_{13}^2)\frac{x^2}{2}-(a_{12}^2+a_{23}^2)\frac{y^2}{2}-(a_{13}^2+a_{23}^2)\frac{z^2}{2}\nonumber\\
     &&-a_{13}a_{32}xy-a_{12}a_{23}xz-a_{21}a_{13}yz)
\end{eqnarray} modulo the transformation in \eqref{T5}.

\begin{prop}  Let $A$ be a $3\times 3$ constant  skew-symmetric matrix
and let $B$ be a symmetric matrix function with zero trace
determined by \eqref{2.18}. Then we have the following solutions of
the MHD equations \eqref{M1}-\eqref{M4}:
\begin{equation}\label{2.21}
     \textbf{v}=A\begin{pmatrix}
                  x\\ y\\ z
                  \end{pmatrix},\ \ \ \
     \textbf{H}=B\begin{pmatrix}
                  x\\ y\\ z
                  \end{pmatrix}
\end{equation} and $p$ is given by \eqref{2.20}.
\end{prop}\psp

{\it Case 4: $A=-A^t$, $B=-B^t$.}\psp

Similarly, we set
\begin{equation}\label{2.22}
   N=\begin{pmatrix}
        0        &   a_{23}   &    -a_{13}\\
        -a_{23}  &   0        &    a_{12}\\
        a_{13}   &   -a_{12}  &    0
     \end{pmatrix}
\end{equation} and
\begin{equation}\label{2.23}
(b_{12},b_{13},b_{23})^t=\exp(Nt)\textbf{c},
\end{equation} where $\textbf{c}$ is a three-dimensional constant
vector. Then we get the following proposition:

\begin{prop} Let $A$ be a $3\times 3$ constant  skew-symmetric matrix
and let $B$ be a  skew-symmetric matrix function determined by
\eqref{2.23}. Then we have the following solutions of the MHD
equations \eqref{M1}-\eqref{M4}:
\begin{equation}\label{2.24}
     \textbf{v}=A\begin{pmatrix}
                  x\\ y\\ z
                  \end{pmatrix},\ \ \ \
     \textbf{H}=B\begin{pmatrix}
                  x\\ y\\ z
                  \end{pmatrix}
\end{equation} and
\begin{eqnarray}\label{2.25}
p&=&-\rho((-(a_{12}^2+a_{13}^2)+2\mu_0(b_{12}^2+b_{13}^2))\frac{x^2}{2}+(-(a_{12}^2+a_{23}^2)+2\mu_0(b_{12}^2+b_{23}^2))\frac{y^2}{2}\nonumber\\
 &&+(-(a_{13}^2+a_{23}^2)+2\mu_0(b_{13}^2+b_{23}^2))\frac{z^2}{2}+(-a_{13}a_{32}+2\mu_0b_{13}b_{32})xy\nonumber\\
 &&+(-a_{12}a_{23}+2\mu_0b_{12}b_{23})xz+(-a_{21}a_{13}+2\mu_0b_{21}b_{13})yz).
\end{eqnarray}
\end{prop}
\section{Asymmetric Approaches}

In this section, we will solve the MHD equations
\eqref{M1}-\eqref{M4} by imposing asymmetric assumptions on
$\textbf{v}$ and $\textbf{H}$, following \cite{X1}.

Firstly, we set
\begin{equation}\label{3.1}
 \left\{\begin{aligned}
          u&=f+(\beta-\psi_z)x,\\
          v&=-\beta y,\\
          w&=\psi
        \end{aligned}
 \right.\ \ \textrm{and}\ \
 \left\{\begin{aligned}
          H^1&=g+(\gamma-\varphi_z)x,\\
          H^2&=-\gamma y,\\
          H^3&=\varphi,
        \end{aligned}
  \right.
\end{equation}where $f$ and $g$ are functions of $t,\ y$ and $z$;
 $\varphi$ and $\psi$ are functions of $t$ and $z$; $\beta$ and
$\gamma$ are functions of $t$.

By \eqref{2.1}-\eqref{2.3},
\begin{eqnarray}\label{3.2}
   W_1&=&f_t+f(\beta-\psi_z)-\beta yf_y+\psi f_z-\nu(f_{yy}+f_{zz})-\mu_0(\gamma yg_y -\varphi g_z)\nonumber\\
      &&+((\beta-\psi_z)^2+\beta'-\psi_{zt}-\psi\psi_{zz}+\nu\psi_{zzz}-\mu_0\varphi\varphi_{zz})x,
\end{eqnarray}
\begin{equation}\label{3.3}
   W_2=(\beta^2+\beta')y-\mu_0g_yg-\mu_0g_y(\gamma-\varphi_z)x,
\end{equation}
and
\begin{eqnarray}\label{3.4}
   W_3=\psi_z(1+\psi)-\nu\psi_{zz}-\mu_0(g_zg+(g_z(\gamma-\varphi_z)-g\varphi_{zz})x-\varphi_{zz}(\gamma-\varphi_z)x^2).
\end{eqnarray}
According to \eqref{M4} and \eqref{2.4},  we find that $\gamma=b/2$
is a constant and
\begin{equation}\label{3.5}
    \left\{\begin{aligned}
               \varphi_t&=-\psi\varphi_z+\psi_z\varphi+\eta\varphi_{zz},\\
               g_t&=-\frac{b}{2}(yf_y-f)+\beta(g+yg_y)-\psi_zg-\psi
               g_z+f\varphi_z+f_z\varphi+\eta(g_{yy}+g_{zz}),
           \end{aligned}
    \right.
\end{equation}
\begin{equation}\label{3.6}
   \left\{\begin{aligned}
              &f_{yt}-\psi_zf_y-\beta yf_{yy}+\psi
                  f_{yz}-\nu(f_{yyy}+f_{zzy})-\mu_0(\frac{b}{2}yg_{yy}+g_y\varphi_z-\varphi
                  g_{yz})=0,\\
              &f_{zt}+f_z\beta-f\psi_{zz}-\beta yf_{yz}+\psi
              f_{zz}-\nu(f_{yyz}+f_{zzz})-\mu_0(\frac{b}{2}yg_{yz}-\varphi
              g_{zz}-\frac{b}{2}g_z+g\varphi_{zz})=0,
          \end{aligned}
    \right.
\end{equation}
\begin{equation}\label{3.7}
   -2\beta\psi_{zz}+\psi_z\psi_{zz}-\psi_{zzt}-\psi\psi_{zzz}+\nu\psi_{zzzz}-\mu_0(b-\varphi_z\varphi_{zz}+\varphi\varphi_{zzz})=0.
\end{equation}
One shows that the system \eqref{3.6} is implied by
\begin{equation}\label{3.8}
   f_t+(\beta-\psi_z)f-\beta yf_y+\psi
   f_z-\nu(f_{yy}+f_{zz})-\mu_0(\frac{b}{2}yg_y-\varphi
   g_z+g(\frac{b}{2}-\varphi_z))=0.
\end{equation}

We assume that
\begin{equation}\label{3.9}
   \varphi=az+l.
\end{equation} Then
\begin{equation}\label{3.10}
\psi=ahz+hl-\frac{l'}{a}
\end{equation} by $\eqref{3.5}_1$, where $a$ is a constant and $h,\;l$ are functions of
$t$. Hence the equations \eqref{3.5}-\eqref{3.7} can be written as
\begin{equation}\label{3.11}
   \left\{\begin{aligned}
       &f_t+(\beta-ah)f-\beta
          yf_y+(h(az+l)-\frac{l'}{a})f_z
          -\mu_0(\frac{b}{2}yg_y-(az+l)g_z-(\frac{b}{2}-a)g)\\&-\nu(f_{yy}+f_{zz})=0,\\
       &g_t+\frac{b}{2}(yf_y+f)-(\beta-ah)g-\beta
       yg_y+(ahz+hl-\frac{l'}{a})g_z-af-(az+l)f_z-\eta(g_{yy}+g_{zz})=0.
   \end{aligned}\right.
\end{equation}
Now we solve above equations. Firstly we suppose that $l$ is a
constant and then use the method of separation of variables. Set
\begin{equation}\label{3.12}
f=y(az+l)s,\ \ \ \ g=y(az+l)q,
\end{equation} where $s$ and $q$ are functions of $t$. Substituting
\eqref{3.12} into \eqref{3.11}, one gets that $s$ is a constant and
\begin{equation}\label{3.13}
q=\exp(2\int(\beta-ah)\md t)((2a-b)s\int\exp(-2\int(\beta-ah)\md
t)\md t+c),
\end{equation}where $c$ is a constant. Furthermore,
\begin{eqnarray}\label{3.14}
p&=&-\rho((\beta'-ah'+(\beta-ah)^2)\frac{x^2}{2}+(-\beta'+\beta^2)\frac{y^2}{2}+\frac{1}{2a}(h'+ah^2)a^2z^2\nonumber\\
 &&-\mu_0q^2a^2z^2\frac{y^2}{2}-\mu_0qa(\frac{b}{2}-a)xyz)
\end{eqnarray}modulo the transformations in \eqref{T4} and in \eqref{T5}.

\begin{thm} For any functions $h(t)$, $\beta(t)$ and constants $a$,
$b$, $c$, and $s$, we have the following solutions of the MHD
equations \eqref{M1}-\eqref{M4}:
\begin{equation}\label{3.15}
 \left\{\begin{aligned}
          u&=asyz+(\beta-ah)x,\\
          v&=-\beta y,\\
          w&=ahz,
        \end{aligned}
 \right.
 \ \ \ \ \ \ \ \ \ \ \ \ \ \ \ \ \ \ \ \ \ \ \ \ \ \ \ \ \ \ \ \ \ \ \ \ \ \ \ \ \ \ \ \ \ \ \ \ \ \ \ \ \ \ \ \ \ \ \ \ \ \ \ \ \ \
 \end{equation}
 \begin{equation}\label{3.16}
 \left\{\begin{aligned}
          H^1&=ayz\exp(2\int(\beta-ah)\md t)((2a-b)s\int\exp(-2\int(\beta-ah)\md
t)\md t+c)+(\frac{b}{2}-a)x,\\
          H^2&=-\frac{b}{2}y,\\
          H^3&=az
        \end{aligned}
  \right.
\end{equation} and $p$ is given by \eqref{3.14}.
\end{thm}

Now we give another solutions of \eqref{3.11}. Motivated from
\eqref{3.12}, we assume that
\begin{equation}\label{3.17}
\left\{\begin{aligned}
    f&=Cy^2+Dz^2+Eyz+Fy+Gz+H,\\
    g&=Iy^2+Jz^2+Kyz+Ly+Mz+N,
 \end{aligned}
  \right.
\end{equation} where $C$, $D$, $E$, $F$, $G$, $H$, $I$, $J$, $K$,
$L$, $M$ and $N$ are functions of $t$.

Substituting \eqref{3.17} into \eqref{3.11}, we get that
\begin{equation}\label{3.18}
\left\{\begin{aligned}
     &C'-(\beta+ah)C-\mu_0(\frac{b}{2}+a)I=0,\\
     &D'+(\beta+ah)D+\mu_0(\frac{b}{2}+a)J=0,\\
     &E'=0,\\
     &F'-ahF+(hl-\frac{l'}{a})E-a\mu_0L+\mu_0lK=0,\\
     &G'+\beta G+2(hl-\frac{l'}{a})D+\frac{b}{2}\mu_0M+2\mu_0lJ=0,\\
     &H'+(\beta-ah)H+(hl-\frac{l'}{a})G-2\nu(C+E)+\mu_0lM+\mu_0(\frac{b}{2}-a)N=0
\end{aligned}
  \right.
\end{equation}
and
\begin{equation}\label{3.19}
\left\{\begin{aligned}
     &I'-(3\beta-ah)I+(\frac{3b}{2}-a)C=0,\\
     &J'-(\beta-3ah)J+(\frac{b}{2}-3a)D=0,\\
     &K'-2(\beta-ah)K+2(b-2a)E=0,\\
     &L'-(2\beta-ah)L+(b-a)F+(hl-\frac{l'}{a})K-lE=0,\\
     &M'-(\beta-2ah)M+(\frac{b}{2}-2a)G+2(hl-\frac{l'}{a})J-2lD=0,\\
     &N'-(\beta-ah)N+(\frac{b}{2}-a)H+(hl-\frac{l'}{a})M-2\eta(I+J)-lG=0.
\end{aligned}
  \right.
\end{equation}
Thus we have that
\begin{equation}\label{3.20}
\left\{\begin{aligned}
     &I=\frac{1}{(\frac{b}{2}+a)\mu_0}(C'-(\beta+ah)C),\\
     &C''-4\beta C'-(\beta'+ah'-(3\beta-ah)(\beta+ah)-(\frac{3b}{2}-a)(\frac{b}{2}+a)\mu_0)C=0,
\end{aligned}
  \right.
\end{equation}
\begin{equation}\label{3.21}
\left\{\begin{aligned}
     &J=-\frac{1}{(\frac{b}{2}+a)\mu_0}(D'+(\beta+ah)D),\\
     &D''+4ah D'+(\beta'+ah'-(\beta-3ah)(\beta+ah)-(\frac{b}{2}-3a)(\frac{b}{2}+a)\mu_0)D=0,
\end{aligned}
  \right.
\end{equation}
\begin{equation}\label{3.22}
\left\{\begin{aligned}
     &E'=0,\\
     &K'-2(\beta-ah)K=-(b-2a)E,
\end{aligned}
  \right.
\end{equation}
\begin{equation}\label{3.23}
\left\{\begin{aligned}
     &L=\frac{1}{a\mu_0}(F'-ahF+(hl-\frac{l'}{a})E+\mu_0lK),\\
     &F''-2\beta F'+(ah(2\beta-ah)-ah'+(b-a)a\mu_0)F=f_F(t),
\end{aligned}
  \right.
\end{equation}
\begin{equation}\label{3.24}
\left\{\begin{aligned}
     &M=-\frac{2}{b\mu_0}(G'+\beta G+2(hl-\frac{l'}{a})D+2\mu_0lJ),\\
     &G''+2ahG'+(\beta'-\beta(\beta-2ah)-\frac{b}{2}(\frac{b}{2}-2a)\mu_0)G=f_G(t)
\end{aligned}
  \right.
\end{equation}
and
\begin{equation}\label{3.25}
\left\{\begin{aligned}
     &N=-\frac{1}{(\frac{b}{2}-a)\mu_0}(H'+(\beta-ah)H+(hl-\frac{l'}{a})G-2\nu(C+E)+\mu_0lM),\\
     &H''+(\beta'-ah'-(\beta-ah)^2-(\frac{b}{2}-a)^2\mu_0)H=f_H(t),
\end{aligned}
  \right.
\end{equation}
where
\begin{equation}\label{3.26}
       f_F(t)=-((hl-\frac{l'}{a})E+\mu_0lK)'+2(\beta-ah)((hl-\frac{l'}{a})E+\mu_0lK)-a\mu_0((hl-\frac{l'}{a})K-lE),
\end{equation}
\begin{equation}\label{3.27}
       f_G(t)=-2((hl-\frac{l'}{a})D+\mu_0lJ)'+2(\beta-2ah)((hl-\frac{l'}{a})D+\mu_0lJ)+b\mu_0((hl-\frac{l'}{a})J-lD),
\end{equation} and
\begin{eqnarray}\label{3.28}
       f_H(t)&=&(-(hl-\frac{l'}{a})G+2\nu(C+E)-\mu_0lM)'+(\beta-ah)((hl-\frac{l'}{a})G-2\nu(C+E)+\mu_0lM)\nonumber\\
              &&+(\frac{b}{2}-a)\mu_0((hl-\frac{l'}{a})M-2\eta(I+J)-lG).
\end{eqnarray}
To write down the solutions of above systems explicitly, we assume
\begin{equation}\label{3.29}
    \beta-ah=q,
\end{equation} where $q$ is a constant. Under this condition, the
equations $\eqref{3.20}_2$, $\eqref{3.21}_2$, $\eqref{3.23}_2$ and
$\eqref{3.24}_2$ can be written as the following form via certain
changing of variables:
\begin{equation}\label{3.30}
A''(t)+kA(t)=f_A(t).
\end{equation} Here $k$ is a constant. We will explain this by solving $\eqref{3.20}_2$ in
detail. Since the other equations are similar, we will only write
down the solutions but omit the solving procedure.

Set
\begin{equation}\label{3.31}
C=\exp(2\int\beta\md t)\tilde{C}.
\end{equation}
Then $\eqref{3.20}_2$ becomes
\begin{equation}\label{3.32}
\tilde{C}''-(q^2-(\frac{3b}{2}-a)(\frac{b}{2}+a)\mu_0)\tilde{C}=0.
\end{equation}
We write $\lambda_C=q^2-(\frac{3b}{2}-a)(\frac{b}{2}+a)\mu_0$ and
define
\begin{equation}\label{3.33}
    \xi_{C_1}=\left\{ \begin{array}{ll}
                         \exp(\sqrt{\lambda_C}\ t+2\int\beta\md t) & \textrm{if }\lambda_C>0,\\
                         \exp(2\int\beta\md t)& \textrm{if }\lambda_C=0,\\
                         \cos(i\sqrt{\lambda_C}\ t)\exp(2\int\beta\md t)& \textrm{if
                         }\lambda_C<0,
                      \end{array}\right.
\end{equation}
\begin{equation}\label{3.34}
    \xi_{C_2}=\left\{ \begin{array}{ll}
                         \exp(-\sqrt{\lambda_C}\ t+2\int\beta\md t) & \textrm{if }\lambda_C>0,\\
                         t\exp(2\int\beta\md t)& \textrm{if }\lambda_C=0,\\
                         \sin(i\sqrt{\lambda_C}\ t)\exp(2\int\beta\md t)& \textrm{if
                         }\lambda_C<0.
                      \end{array}\right.
\end{equation}
Then
\begin{equation}\label{3.35}
C=l_{C_1}\xi_{C_1}+l_{C_2}\xi_{C_2}
\end{equation} and
\begin{equation}\label{3.36}
I=\frac{\sum\limits_{j=1}^2l_{C_j}(\xi_{C_j}'-(2\beta-q)\xi_{C_j})}{(\frac{b}{2}+a)\mu_0},
\end{equation}
where $l_{C_1}$ and $l_{C_2}$ are real constants.

Similarly, we set
$\lambda_D=q^2+(\frac{b}{2}-3a)(\frac{b}{2}+a)\mu_0$ and define
\begin{equation}\label{3.37}
     \xi_{D_1}=\left\{ \begin{array}{ll}
                            \exp(\sqrt{\lambda_D}\ t-2\int(\beta+q)\md t) & \textrm{if }\lambda_D>0,\\
                            \exp(-2\int(\beta+q)\md t)& \textrm{if }\lambda_D=0,\\
                            \cos(i\sqrt{\lambda_D}\ t)\exp(-2\int(\beta+q)\md t)& \textrm{if
                              }\lambda_D<0,
                       \end{array}\right.
\end{equation}
\begin{equation}\label{3.38}
     \xi_{D_2}=\left\{ \begin{array}{ll}
                            \exp(-\sqrt{\lambda_D}\ t-2\int(\beta+q)\md t) & \textrm{if }\lambda_D>0,\\
                            t\exp(-2\int(\beta+q)\md t)& \textrm{if }\lambda_D=0,\\
                            \sin(i\sqrt{\lambda_D}\ t)\exp(-2\int(\beta+q)\md t)& \textrm{if
                              }\lambda_D<0.
                       \end{array}\right.
\end{equation}
Then
\begin{equation}\label{3.39}
D=l_{D_1}\xi_{D_1}+l_{D_2}\xi_{D_2}
\end{equation} and
\begin{equation}\label{3.40}
J=-\frac{\sum\limits_{j=1}^2l_{D_j}(\xi_{D_j}'+(2\beta-q)\xi_{D_j})}{(\frac{b}{2}+a)\mu_0},
\end{equation}
where $l_{D_1}$ and $l_{D_2}$ are real constants.

Given functions $r(t)$ and $s(t)$, we denote their Wronskian
determinant by
\begin{equation}\label{3.41}
    W[r(t),s(t)]=\det\begin{pmatrix}
                           r(t) &  s(t)\\
                           r'(t)&  s'(t)
    \end{pmatrix}=r(t)s'(t)-r'(t)s(t).
\end{equation} For notational convenience, we write
\begin{equation}\label{3.42}
     W_1[r(t),s(t)]=\det\begin{pmatrix}
                           0 &  s(t)\\
                           1&  s'(t)
                       \end{pmatrix}=-s(t),
\end{equation}
\begin{equation}\label{3.43}
     W_2[r(t),s(t)]=\det\begin{pmatrix}
                            r(t) &  0\\
                           r'(t)&  1
                       \end{pmatrix}=r(t).\end{equation}

Set $\lambda_F=q^2-(b-a)a\mu_0$ and define
\begin{equation}\label{3.44}
     \xi_{F_1}=\left\{ \begin{array}{ll}
                            \exp(\sqrt{\lambda_F}\ t+\int\beta\md t) & \textrm{if }\lambda_F>0,\\
                            \exp(\int\beta\md t)& \textrm{if }\lambda_F=0,\\
                            \cos(i\sqrt{\lambda_F}\ t)\exp(\int\beta\md t)& \textrm{if
                              }\lambda_F<0,
                       \end{array}\right.
\end{equation}
\begin{equation}\label{3.45}
     \xi_{F_2}=\left\{ \begin{array}{ll}
                            \exp(-\sqrt{\lambda_F}\ t+\int\beta\md t) & \textrm{if }\lambda_F>0,\\
                            t\exp(\int\beta\md t)& \textrm{if }\lambda_F=0,\\
                            \sin(i\sqrt{\lambda_F}\ t)\exp(\int\beta\md t)& \textrm{if
                              }\lambda_F<0.
                       \end{array}\right.
\end{equation}
Then
\begin{equation}\label{3.46}
F=\sum_{j=1}^2\xi_{F_j}(l_{F_j}+\int_0^t\frac{W_j[\xi_{F_1}(s),\xi_{F_2}(s)]}{W[\xi_{F_1}(s),\xi_{F_2}(s)]}f_F(s)\md
s)
\end{equation}
and
\begin{eqnarray}\label{3.47}
L&=&\frac{1}{a\mu_0}\Big(\frac{1}{a}((\beta-q)l-l'+\frac{a(b-2a)}{2q})E+ke^{2qt}+\sum_{j=1}^2(\xi_{F_j}
\frac{W_j[\xi_{F_1}(t),\xi_{F_2}(t)]}{W[\xi_{F_1}(t),\xi_{F_2}(t)]}f_F(t)\nonumber\\
  &&+(\xi_{F_j}'-(\beta-q)\xi_{F_j})(l_{F_j}+\int_0^t\frac{W_j[\xi_{F_1}(s),\xi_{F_2}(s)]}{W[\xi_{F_1}(s),\xi_{F_2}(s)]}f_F(s)\md
s))\Big),
\end{eqnarray} where $l_{F_1}$ and $l_{F_2}$ are real constants.
In this expression, we have use the fact:
\begin{equation}\label{3.48}
K=ke^{2qt}+\frac{b-2a}{2q}E,
\end{equation} where $k$ and $E$ are real constant.

Let $\lambda_G=q^2+\frac{b}{2}(\frac{b}{2}-2a)\mu_0$ and denote
\begin{equation}\label{3.49}
   \xi_{G_1}=\left\{ \begin{array}{ll}
                            \exp(\sqrt{\lambda_G}\ t-\int(\beta-q)\md t) & \textrm{if }\lambda_G>0,\\
                            \exp(-\int(\beta-q)\md t)& \textrm{if }\lambda_G=0,\\
                            \cos(i\sqrt{\lambda_G}\ t)\exp(-\int(\beta-q)\md t)& \textrm{if
                              }\lambda_G<0,
                       \end{array}\right.
\end{equation}
\begin{equation}\label{3.50}
   \xi_{G_2}=\left\{ \begin{array}{ll}
                            \exp(-\sqrt{\lambda_G}\ t-\int(\beta-q)\md t) & \textrm{if }\lambda_G>0,\\
                            t\exp(-\int(\beta-q)\md t)& \textrm{if }\lambda_G=0,\\
                            \sin(i\sqrt{\lambda_G}\ t)\exp(-\int(\beta-q)\md t)& \textrm{if
                              }\lambda_G<0.
                       \end{array}\right.
\end{equation}
Then
\begin{equation}\label{3.51}
G=\sum_{j=1}^2\xi_{G_j}(l_{G_j}+\int_0^t\frac{W_j[\xi_{G_1}(s),\xi_{G_2}(s)]}{W[\xi_{G_1}(s),\xi_{G_2}(s)]}f_G(s)\md
s)
\end{equation}
and
\begin{eqnarray}\label{3.52}
M&=&-\frac{2}{\mu_0}\Big(\sum_{j=1}^2(2l_{D_j}(\frac{(\beta-q-l')\xi_{D_j}}{a}-\frac{l}{\frac{b}{2}+a}(\xi_{D_j}'+(2\beta-q)\xi_{D_j}))\nonumber\\
  &&+(\xi_{G_j}'+\beta\xi_{G_j})(l_{G_j}+\int_0^t\frac{W_j[\xi_{G_1}(s),\xi_{G_2}(s)]}{W[\xi_{G_1}(s),\xi_{G_2}(s)]}f_G(s)\md s)\nonumber\\
  &&+\xi_{G_j}\frac{W_j[\xi_{G_1}(t),\xi_{G_2}(t)]}{W[\xi_{G_1}(t),\xi_{G_2}(t)]}f_G(t))\Big),
\end{eqnarray}
where $l_{G_1}$ and $l_{G_2}$ are real constants.

Set $\lambda_H=q^2+(\frac{b}{2}-a)\mu_0^2$ and define
\begin{equation}\label{3.53}
    \xi_{H_1}=\left\{ \begin{array}{ll}
                            \exp(\sqrt{\lambda_H}\ t) & \textrm{if }\lambda_H>0,\\
                            1& \textrm{if }\lambda_H=0,\\
                            \cos(i\sqrt{\lambda_H}\ t)& \textrm{if
                              }\lambda_H<0,
                       \end{array}\right.
\end{equation}
\begin{equation}\label{3.54}
    \xi_{H_2}=\left\{ \begin{array}{ll}
                            \exp(-\sqrt{\lambda_H}\ t) & \textrm{if }\lambda_H>0,\\
                            t& \textrm{if }\lambda_H=0,\\
                            \sin(i\sqrt{\lambda_H}\ t)& \textrm{if
                              }\lambda_H<0.
                       \end{array}\right.
\end{equation}
Then
\begin{equation}\label{3.55}
H=\sum_{j=1}^2\xi_{H_j}(l_{H_j}+\int_0^t\frac{W_j[\xi_{H_1}(s),\xi_{H_2}(s)]}{W[\xi_{H_1}(s),\xi_{H_2}(s)]}f_H(s)\md
s)
\end{equation} and
\begin{eqnarray}\label{3.56}
N&=&-\frac{1}{(\frac{b}{2}-a)\mu_0}\Big(\frac{1}{a}((\beta-q)l-l')G-2\nu(C+E)+\mu_0lM+\sum_{j=1}^2(\xi_{H_j}
\frac{W_j[\xi_{H_1}(t),\xi_{H_2}(t)]}{W[\xi_{H_1}(t),\xi_{H_2}(t)]}f_H(t)\nonumber\\
  &&+(\xi_{H_j}'+q\xi_{H_j})(l_{H_j}+\int_0^t\frac{W_j[\xi_{H_1}(s),\xi_{H_2}(s)]}{W[\xi_{H_1}(s),\xi_{H_2}(s)]}f_H(s)\md
s))\Big),
\end{eqnarray}
where $E$, $l_{H_1}$ and $l_{H_2}$ are real constants, $C$, $G$ and
$M$ are given by \eqref{3.35}, \eqref{3.51} and \eqref{3.52}.

Furthermore, by \eqref{2.11}, one has
\begin{eqnarray}\label{3.57}
 p&=&-\rho\Big(\frac{q^2}{2}x^2+(H'+qH+\frac{(\beta-q)l-l'}{a}G+\mu_0lM-2\nu(C+D))x\nonumber\\
     &&-\frac{\mu_0}{2}I^2y^4-\mu_0ILy^3+\frac{1}{2}(-\beta'+\beta^2-\mu_0(2IN+L^2))y^2-\mu_0LNy\nonumber\\
     &&-\frac{\mu_0}{2}J^2z^4-\mu_0JMz^3+(\beta'+(\beta-q)^2-\mu_0(2JN+M^2))z^2\nonumber\\
     &&+(-\mu_0MN+\frac{1}{a}(\beta'l+(\beta-q)l'-l'')+\frac{\beta-q}{a}((\beta-q)l-l'))z\nonumber\\
     &&-(\frac{b}{2}-a)\mu_0Kxyz-\frac{b-2a}{2}\mu_0Ixy^2-(\frac{b}{2}-a)\mu_0Lxy-(\frac{b}{2}-a)\mu_0Jxz^2-(\frac{b}{2}-a)\mu_0Mxz\nonumber\\
     &&-\mu_0KJyz^3-\mu_0KIy^3z-\frac{\mu_0}{2}(2IJ+K^2)y^2z^2-\mu_0(JL+KM)yz^2\nonumber\\
     &&-\mu_0(KL+IM)y^z-\mu_0(KN+LM)yz\Big)
\end{eqnarray} modulo the transformation in \eqref{T5}.

\begin{thm}
For any functions $\beta(t)$, $l(t)$ and real constants $a$, $b$,
$q$, $E$, we have the following solutions of the MHD equations
\eqref{M1}-\eqref{M4}:
\begin{equation}\label{3.58}
 \left\{\begin{aligned}
          u&=Cy^2+Dz^2+Eyz+Fy+Gz+H+qx,\\
          v&=-\beta y,\\
          w&=(\beta-q)z+\frac{(\beta-q)l-l'}{a},
        \end{aligned}
 \right.
\end{equation}
\begin{equation}\label{3.59}
 \left\{\begin{aligned}
          H^1&=Iy^2+Jz^2+Kyz+Ly+Mz+N+(\frac{b}{2}-a)x,\\
          H^2&=-\frac{b}{2}y,\\
          H^3&=az+l
        \end{aligned}
  \right.
\end{equation}and $p$ is given by \eqref{3.57}. Here the functions
$C$, $D$, $F$, $G$, $H$, $I$, $J$, $K$, $L$, $M$ and $N$ are given
by \eqref{3.35}, \eqref{3.39}, \eqref{3.46}, \eqref{3.51},
\eqref{3.55}, \eqref{3.36}, \eqref{3.40}, \eqref{3.48},
\eqref{3.47}, \eqref{3.52} and \eqref{3.56}, respectively.
\end{thm}

\vskip 1.0cm

We denote
\begin{equation}\label{3.60}
\varpi=x^2+y^2
\end{equation} and assume
\begin{equation}\label{3.61}
   \left\{\begin{aligned}
     u&=\frac{\alpha}{2}x+y\phi(t,\varpi),\\
     v&=\frac{\alpha}{2}y-x\phi(t,\varpi),\\
     w&=-\alpha z+\psi(t,\varpi),
   \end{aligned}
   \right.
\ \ \ \ \ \ \ \ \ \ \ \
   \left\{\begin{aligned}
     H^1&=\frac{\beta}{2}x+y\xi(t,\varpi),\\
     H^2&=\frac{\beta}{2}y-x\xi(t,\varpi),\\
     H^3&=-\beta z+\sigma(t,\varpi).
   \end{aligned}
   \right.
\end{equation}
Recall that \eqref{M4} is just
\begin{equation}\label{3.62}
   \left\{\begin{aligned}
     H_t^1&=(uH^2-vH^1)_y-(wH^1-uH^3)_z+\eta\Delta H^1,\\
     H_t^2&=(vH^3-wH^2)_z-(uH^2-vH^1)_x+\eta\Delta H^2,\\
     H_t^3&=(wH^1-uH^3)_x-(vH^3-wH^2)_y+\eta\Delta H^3.
   \end{aligned}\right.
\end{equation}
Substituting \eqref{3.61} into $\eqref{3.62}_1$, one gets that
$\beta=b$ is a constant and
\begin{equation}\label{3.63}
   \xi_t=-\alpha\varpi\xi_{\varpi}+b\varpi\phi_{\varpi}+4\eta(\varpi\xi)_{\varpi\varpi}.
\end{equation} One shows that $\eqref{3.62}_2$ is
equivalent to \eqref{3.63}, and $\eqref{3.62}_3$ implies that
\begin{equation}\label{3.64}
  \sigma_t=b(\varpi\psi)_{\varpi}-\alpha\varpi\sigma_{\varpi}+4\eta(\sigma_{\varpi}+\varpi\sigma_{\varpi\varpi}).
\end{equation} Moreover, by \eqref{2.4}, we have
\begin{equation}\label{3.65}
  \left\{\begin{aligned}
    (\varpi\phi)_{t\varpi}+\alpha(\varpi(\varpi\phi)_{\varpi})_{\varpi}-4\nu(\varpi(\varpi\phi)_{\varpi\varpi})_{\varpi}
              +b\mu_0(\varpi(\varpi\xi)_{\varpi})_{\varpi}&=0,\\
    \psi_t-\alpha\psi+\alpha\varpi\psi_{\varpi}-4\nu(\psi_{\varpi}+\varpi\psi_{\varpi\varpi})+b\mu_0\varpi\sigma_{\varpi}-b\mu_0\sigma&=0.
  \end{aligned}
  \right.
\end{equation}
Given any $n\in\mathbb{N}$, set
\begin{equation}\label{3.66}
\phi=\sum_{m=0}^n a_m(t)\varpi^m,\ \ \ \ \xi=\sum_{m=0}^n
c_m(t)\varpi^m,\ \ \ \ \sigma=\sum_{m=0}^nf_m(t)\varpi^m,\ \ \ \
\psi=\sum_{m=0}^n g_m(t)\varpi^m.
\end{equation}
Substituting \eqref{3.66} into \eqref{3.63}-\eqref{3.65} and
considering the coefficients of $\varpi^m$, we obtain that
\begin{equation}\label{3.67}
 \left\{\begin{aligned}
   &a_{m,t}+(m+1)\alpha a_m+b\mu_0(m+1)c_m=4\nu(m+1)(m+2)a_{m+1},\\
   &c_{m,t}-bma_m+\alpha mc_m=4\eta(m+1)(m+2)c_{m+1}
 \end{aligned}\right.
\end{equation} and
\begin{equation}\label{3.68}
\left\{\begin{aligned}
   f_{m,t}&=-\alpha mf_m+b(m+1)g_m+4\eta(m+1)^2f_{m+1},\\
   g_{m,t}&=-(m-1)\alpha g_m-b(m-1)\mu_0f_m+4\nu(m+1)^2g_{m+1},
 \end{aligned}\right.
\end{equation}where $a_{-k}=a_{n+k}=c_{-k}=c_{n+k}=f_{-k}=f_{n+k}=g_{-k}=g_{n+k}=0$ for $k>0$. These systems are similar to \eqref{3.20}. If we set
\begin{equation}\label{3.69}
  \alpha=\frac{k+2k^2le^{-kt}}{-1+2kle^{-kt}}
\end{equation} for any constants $k$ and $l$, then we can transform
them into certain second order ordinary differential equations with
constant coefficients, as we did in \eqref{3.20}.

Observe that
\begin{equation}\label{3.70}
c_m=-\frac{1}{(m+1)b\mu_0}(a_{m,t}+(m+1)\alpha
a_m-4\nu(m+1)(m+2)a_{m+1}).
\end{equation} Substituting \eqref{3.70} into \eqref{3.67}, one has
\begin{equation}\label{3.71}
a_m''+(2m+1)\alpha
a_m'+(m+1)(m\alpha^2+\alpha'+b^2\mu_0m)a_m=F_m(t),
\end{equation}
where
\begin{equation}\label{3.72}
F_m=4(m+1)(m+2)(m\nu\alpha a_{m+1}+\nu a_{m+1}'+\eta c_{m+1}).
\end{equation}
Set
\begin{equation}\label{3.73}
a_m=\tilde{a}_m\exp(-\frac{1}{2}\int(2m+1)\alpha\md
t)=(e^{(m+\frac{1}{2})kt}-2kle^{(m-\frac{1}{2})kt})\tilde{a}_m.
\end{equation} Then \eqref{3.71} becomes
\begin{equation}\label{3.74}
\tilde{a}_m''+((m+1)(m\alpha^2+\alpha'+b^2\mu_0m)-\frac{2m+1}{2}\alpha'-\frac{1}{4}(2m+1)^2\alpha^2)\tilde{a}_m=F_m(t).
\end{equation}
By \eqref{3.69},
\begin{equation}\label{3.75}
 (m+1)(m\alpha^2+\alpha'+b^2\mu_0m)-\frac{2m+1}{2}\alpha'-\frac{1}{4}(2m+1)^2\alpha^2=m(m+1)b^2\mu_0-\frac{1}{4}k^2
\end{equation} is a constant. Let
$\lambda_m=-m(m+1)b^2\mu_0+\frac{1}{4}k^2$ and define
\begin{equation}\label{3.76}
 \zeta_{m,1}=\left\{ \begin{array}{ll}
                            \exp(\sqrt{\lambda_m}\ t)(e^{(m+\frac{1}{2})kt}-2kle^{(m-\frac{1}{2})kt}) & \textrm{if }\lambda_m>0,\\
                            e^{(m+\frac{1}{2})kt}-2kle^{(m-\frac{1}{2})kt}& \textrm{if }\lambda_m=0,\\
                            \cos(i\sqrt{\lambda_m}\ t)(e^{(m+\frac{1}{2})kt}-2kle^{(m-\frac{1}{2})kt})& \textrm{if
                              }\lambda_m<0,
                       \end{array}\right.
\end{equation}
\begin{equation}\label{3.77}
 \zeta_{m,2}=\left\{ \begin{array}{ll}
                            \exp(-\sqrt{\lambda_m}\ t)(e^{(m+\frac{1}{2})kt}-2kle^{(m-\frac{1}{2})kt}) & \textrm{if }\lambda_m>0,\\
                            t(e^{(m+\frac{1}{2})kt}-2kle^{(m-\frac{1}{2})kt})& \textrm{if }\lambda_m=0,\\
                            \sin(i\sqrt{\lambda_m}\ t)(e^{(m+\frac{1}{2})kt}-2kle^{(m-\frac{1}{2})kt})& \textrm{if
                              }\lambda_m<0.
                       \end{array}\right.
\end{equation}
Then we can get $a_m$ and $c_m$ by the following recursive formulae
\begin{equation}\label{3.78}
   a_m=\sum_{j=1}^2\zeta_{m,j}(l_{m,j}+\int_0^t\frac{W_j[\zeta_{m,1}(s),\zeta_{m,2}(s)]}{W[\zeta_{m,1}(s),\zeta_{m,2}(s)]}F_m(s)\md
s),
\end{equation}
\begin{eqnarray}\label{3.79}
 c_m&=&-\frac{1}{b\mu_0(m+1)}\sum_{j=1}^2\Big(
         \zeta_{m,j}'(l_{m,j}+\int_0^t\frac{W_j[\zeta_{m,1}(s),\zeta_{m,2}(s)]}{W[\zeta_{m,1}(s),\zeta_{m,2}(s)]}F_m(s)\md
         s)\nonumber\\
    &&+\zeta_{m,j}(\frac{W_j[\zeta_{m,1}(t),\zeta_{m,2}(t)]}{W[\zeta_{m,1}(t),\zeta_{m,2}(t)]}F_m(t)+
         (m+1)\alpha(l_{m,j}+\int_0^t\frac{W_j[\zeta_{m,1}(s),\zeta_{m,2}(s)]}{W[\zeta_{m,1}(s),\zeta_{m,2}(s)]}F_m(s)\md
         s))\nonumber\\
    &&+4\nu(m+1)(m+2)\zeta_{m+1,j}(l_{m+1,j}+\int_0^t\frac{W_j[\zeta_{m+1,1}(s),\zeta_{m+1,2}(s)]}{W[\zeta_{m+1,1}(s),\zeta_{m+1,2}(s)]}F_{m+1}(s)\md
s)\Big),
\end{eqnarray} where $l_{m,1}$ and $l_{m,2}$ are real constants.

Given
\begin{equation}\label{3.80}
  G_m(t)=4(m+1)^2((m-1)\alpha\eta f_{m+1}+\eta f_{m+1}'+\nu
  g_{m+1}),
\end{equation} the solutions of system \eqref{3.68} can be written down
similarly as those of \eqref{3.78} and \eqref{3.79}.

Set
\begin{equation}\label{3.81}
  \tau_m=\frac{1}{4}k^2-(m^2-1)b^2\mu_0
\end{equation} and denote
\begin{equation}\label{3.82}
 \varsigma_{m,1}=\left\{ \begin{array}{ll}
                            \exp(\sqrt{\tau_m}\ t)(e^{(m+\frac{1}{2})kt}-2kle^{(m-\frac{1}{2})kt}) & \textrm{if }\tau_m>0,\\
                            e^{(m+\frac{1}{2})kt}-2kle^{(m-\frac{1}{2})kt}& \textrm{if }\tau_m=0,\\
                            \cos(i\sqrt{\tau_m}\ t)(e^{(m+\frac{1}{2})kt}-2kle^{(m-\frac{1}{2})kt})& \textrm{if
                              }\tau_m<0,
                       \end{array}\right.
\end{equation}
\begin{equation}\label{3.83}
 \varsigma_{m,2}=\left\{ \begin{array}{ll}
                            \exp(-\sqrt{\tau_m}\ t)(e^{(m+\frac{1}{2})kt}-2kle^{(m-\frac{1}{2})kt}) & \textrm{if }\tau_m>0,\\
                            t(e^{(m+\frac{1}{2})kt}-2kle^{(m-\frac{1}{2})kt})& \textrm{if }\tau_m=0,\\
                            \sin(i\sqrt{\tau_m}\ t)(e^{(m+\frac{1}{2})kt}-2kle^{(m-\frac{1}{2})kt})& \textrm{if
                              }\tau_m<0.
                       \end{array}\right.
\end{equation} Then
\begin{equation}\label{3.84}
   f_m=\sum_{j=1}^2\varsigma_{m,j}(q_{m,j}+\int_0^t\frac{W_j[\varsigma_{m,1}(s),\varsigma_{m,2}(s)]}{W[\varsigma_{m,1}(s),\varsigma_{m,2}(s)]}G_m(s)\md
s)
\end{equation} and
\begin{eqnarray}\label{3.85}
 g_m&=&\frac{1}{b(m+1)}\sum_{j=1}^2\Big(
         \varsigma_{m,j}'(q_{m,j}+\int_0^t\frac{W_j[\varsigma_{m,1}(s),\varsigma_{m,2}(s)]}{W[\varsigma_{m,1}(s),\varsigma_{m,2}(s)]}G_m(s)\md
         s)\nonumber\\
    &&+\varsigma_{m,j}(\frac{W_j[\varsigma_{m,1}(t),\varsigma_{m,2}(t)]}{W[\varsigma_{m,1}(t),\varsigma_{m,2}(t)]}G_m(t)+
         m\alpha(q_{m,j}+\int_0^t\frac{W_j[\varsigma_{m,1}(s),\varsigma_{m,2}(s)]}{W[\varsigma_{m,1}(s),\varsigma_{m,2}(s)]}G_m(s)\md
         s))\nonumber\\
    &&-4\eta(m+1)^2\varsigma_{m+1,j}(q_{m+1,j}+\int_0^t\frac{W_j[\varsigma_{m+1,1}(s),\varsigma_{m+1,2}(s)]}
    {W[\varsigma_{m+1,1}(s),\varsigma_{m+1,2}(s)]}G_{m+1}(s)\md
s)\Big),
\end{eqnarray} where $q_{m_1}$ and $q_{m,2}$ are real constants.
The function $p$  can be written as
\begin{equation}\label{3.86}
p=-\rho(\frac{\alpha'+\alpha^2}{2}\varpi+\frac{1}{2}(2\alpha^2-\alpha')z^2+b\mu_0z\sigma-\frac{1}{2}\int\phi^2\md
\varpi-\xi^2+\int\varpi\xi\xi_{\varpi}\md\varpi)
\end{equation}  modulo the transformation in \eqref{T5}.

\begin{thm}
  Given any constant $b$, positive integer $n$ and the function
\begin{equation}\label{3.87}
  \alpha=\frac{k+2k^2le^{-kt}}{-1+2kle^{-kt}}
\end{equation} with constants $k$ and $l$, we have the following
solutions of the MHD equations \eqref{M1}-\eqref{M4}:
\begin{equation}\label{3.88}
   \left\{\begin{aligned}
     u&=\frac{\alpha}{2}x+y\sum_{m=0}^n a_m(t)(x^2+y^2)^m,\\
     v&=\frac{\alpha}{2}y-x\sum_{m=0}^n a_m(t)(x^2+y^2)^m,\\
     w&=-\alpha z+\sum_{m=0}^n g_m(t)(x^2+y^2)^m,
   \end{aligned}
   \right.
\ \ \ \ \ \ \ \ \ \ \ \
   \left\{\begin{aligned}
     H^1&=\frac{b}{2}x+y\sum_{m=0}^nc_m(t)(x^2+y^2)^m,\\
     H^2&=\frac{b}{2}y-x\sum_{m=0}^nc_m(t)(x^2+y^2)^m,\\
     H^3&=-b z+\sum_{m=0}^nf_m(t)(x^2+y^2)^m.
   \end{aligned}
   \right.
\end{equation} and $p$ is given by \eqref{3.86}. Here the functions
$a_m$, $c_m$, $f_m$ and $g_m$ are given by \eqref{3.78},
\eqref{3.79}, \eqref{3.84} and \eqref{3.85}, respectively.
\end{thm}

\section{Moving-Frame Approach}

In this section, we use the moving-frame method to find six families
exact solutions of the MHD equations \eqref{M1}-\eqref{M4}.

Let  $\alpha$ and $\beta$ be two functions in $t$. We denote
\begin{equation}\label{4.1}
   T=\begin{pmatrix}
          \cos\alpha  & \sin\alpha\cos\beta & \sin\alpha\sin\beta\\
          -\sin\alpha & \cos\alpha\cos\beta & \cos\alpha\sin\beta\\
          0           & -\sin\beta          & \cos\beta
     \end{pmatrix}.
\end{equation} Then
\begin{equation}\label{4.2}
   T^{-1}=T^t\begin{pmatrix}
          \cos\alpha  & -\sin\alpha & 0\\
          \sin\alpha\cos\beta & \cos\alpha\cos\beta & -\sin\beta\\
          \sin\alpha\sin\beta &\cos\alpha\sin\beta  & \cos\beta
     \end{pmatrix}.
\end{equation} Following section 3 of \cite{X1}, we define the
moving-frames:
\begin{equation}\label{4.3}
   \widetilde{\textbf{H}}=\begin{pmatrix}
           \h ^1 \\ \h ^2 \\ \h ^3
   \end{pmatrix}
=T\begin{pmatrix}H^1\\H^2\\H^3
   \end{pmatrix},\ \ \ \ \widetilde{\textbf{v}}=\begin{pmatrix}
           \U \\ \V \\ \W
   \end{pmatrix}
=T\begin{pmatrix}u\\v\\w
   \end{pmatrix},\ \ \ \ \begin{pmatrix}
           \x \\ \y \\ \z
   \end{pmatrix}=T\begin{pmatrix}x\\y\\z
   \end{pmatrix}.
\end{equation}
Write
\begin{equation}\label{4.4}
  \widetilde{\nabla}=(\partial_{\x}\ ,\partial_{\y}\ ,\partial_{\z})\
  \ \ \ \textrm{and}\ \ \
  \widetilde{rot}\ \textbf{M}=\widetilde{\nabla}\times\textbf{M}.
\end{equation} Then we have
\begin{lem} We have the following equations:
 \begin{equation}\label{4.5}
         \widetilde{\textbf{H}}\times\widetilde{rot}\ \widetilde{\textbf{H}}=T(\textbf{H}\times rot\
         \textbf{H}),
      \end{equation}

  \begin{equation}\label{4.6}
    \widetilde{rot}(\widetilde{\textbf{v}}\times\widetilde{\textbf{H}})=T(rot\
    (\textbf{v}\times\textbf{H})).
  \end{equation}
\end{lem}
\begin{proof} It follows from straightforward calculation.
\end{proof}\psp

We write
\begin{eqnarray}\label{4.7}
  R_1&=&\U_t+\alpha'(\y\U_{\x}-\x\U_{\y}-\V)+(\z\U_{\x}-\x\U_{\z}-\W)\beta'\sin\alpha\nonumber\\
     &&+(\z\U_{\y}-\y\U_{\z})\beta'\cos\alpha+\U\U_{\x}+\V\U_{\y}+\W\U_{\z}-\nu\Delta(\U),
\end{eqnarray}
\begin{eqnarray}\label{4.8}
   R_2&=&\V_t+\alpha'(\y\V_{\x}-\x\V_{\y}+\U)+(\z\V_{\x}-\x\V_{\z})\beta'\sin\alpha\nonumber\\
       &&+(\z\V_{\y}-\y\V_{\z}-\W)\beta'\cos\alpha+\U\V_{\x}+\V\V_{\y}+\W\V_{\z}-\nu\Delta(\V),
\end{eqnarray}
\begin{eqnarray}\label{4.9}
  R_3&=&\W_t+\alpha'(\y\W_{\x}-\x\W_{\y})+(\z\W_{\x}-\x\W_{\z}+\U)\beta'\sin\alpha\nonumber\\
     &&+(\z\W_{\y}-\y\W_{\z}+\V)\beta'\cos\alpha+\U\W_{\x}+\V\W_{\y}+\W\W_{\z}-\nu\Delta(\W),
\end{eqnarray}
\begin{equation}\label{4.10}
 \left\{\begin{aligned}
   G_1&=\h^2(\h^2_{\x}-\h^1_{\y})-\h^3(\h^1_{\z}-\h^2_{\x}),\\
   G_2&=\h^3(\h^3_{\y}-\h^2_{\z})-\h^1(\h^2_{\x}-\h^3_{\y}),\\
   G_3&=\h^1(\h^1_{\z}-\h^3_{\x})-\h^2(\h^3_{\y}-\h^2_{\z}),
 \end{aligned}
 \right.
\end{equation}
\begin{equation}\label{4.11}
 \Phi_j=R_j-\mu_0G_j\ \ \ \ \ \ \ \ \ \ \ \ \ \ \ \ \ \ \ \ \ j=1,\ 2, \
 3,
 \end{equation}
 \begin{eqnarray}\label{4.12}
  &&\Psi_1=\h^1_t+\alpha'(\y\h^1_{\x}-\x\h^1_{\y}-\h^2)+(\z\h^1_{\x}-\x\h^1_{z}-\h^3)\beta'\sin\alpha\nonumber\\
        &&+(\z\h^1_{\y}-\y\h^1_{\z})\beta'\cos\alpha-(\U\h^2-\V\h^1)_{\y}+(\W\h^1-\U\h^3)_{\z}-\eta\Delta\h^1,
 \end{eqnarray}
\begin{eqnarray}\label{4.13}
&&\Psi_2=\h^2_t+\alpha'(\y\h^2_{\x}-\x\h^2_{\y}+\h^1)+(\z\h^2_{\x}-\x\h^2_{\z})\beta'\sin\alpha\nonumber\\
        &&+(\z\h^2_{\y}-\y\h^2_{\z}-\h^3)\beta'\cos\alpha-(\V\h^3-\W\h^2)_{\z}+(\U\h^2-\V\h^1)_{\x}-\eta\Delta\h^2,\nonumber\\
 \end{eqnarray}
\begin{eqnarray}\label{4.14}
&&\Psi_3=\h^3_t+\alpha'(\y\h^3_{\x}-\x\h^3_{\y})+(\z\h^3_{\x}-\x\h^3_{\z}+\h^1)\beta'\sin\alpha\nonumber\\
        &&+(\z\h^3_{\y}-\y\h^3_{\z}+\h^2)\beta'\cos\alpha-(\W\h^1-\U\h^3)_{\x}+(\V\h^3-\W\h^2)_{\y}-\eta\Delta\h^3.\nonumber\\
 \end{eqnarray}
Then the MHD equations \eqref{M1}-\eqref{M4} become
\begin{equation}\label{4.15}
\Phi_1+\frac{1}{\rho}p_{_{\x}}=0,\ \ \ \ \
\Phi_2+\frac{1}{\rho}p_{_\y}=0,\ \ \ \ \
\Phi_3+\frac{1}{\rho}p_{_\z}=0,
\end{equation}
\begin{equation}\label{4.16}
\Psi_j=0,\ \ \ \ \ \ \ \ \ \ \ \ \ \ \ \ \ \ \ \ \ \ \ \ j=1,\ 2,\
3,
\end{equation}
\begin{equation}\label{4.17}
\left\{ \begin{aligned}
        \U_{\x}+\V_{\y}+\W_{\z}&=0,\\
        \h^1_{\x}+\h^2_{\y}+\h^3_{\z}&=0.
\end{aligned}\right.
\end{equation}

Instead of solving \eqref{4.15}, we will first deal with the
compatibility conditions:
\begin{equation}\label{4.18}
   \partial_{\y}(\Phi_1)=\partial_{\x}(\Phi_2),\ \ \ \ \partial_{\z}(\Phi_1)=\partial_{\x}(\Phi_3),\ \ \ \
   \partial_{\z}(\Phi_2)=\partial_{\y}(\Phi_3).
\end{equation}
Then we find $p$ via \eqref{4.15}.

\vskip 1.0cm

Now we suppose that
\begin{equation}\label{4.26}
   \left\{\begin{aligned}
         \U&=-2\gamma\x-\alpha\y-\z\beta'\sin\alpha,\\
         \V&=\phi+\gamma\y,\\
         \W&=\psi+\gamma\z,
   \end{aligned}\right.\ \ \ \ \textrm{and}\ \ \ \
   \left\{\begin{aligned}
         \h^1&=g-2\xi\x,\\
         \h^2&=\sigma+\xi\y,\\
         \h^3&=h+\xi\z,
   \end{aligned}\right.
\end{equation}
where $\phi$, $\psi$, $\sigma$ and $h$ are functions in $\x$ and
$t$, and $g$, $\gamma$ and $\xi$ are functions in $t$.

Substituting \eqref{4.26} into \eqref{4.16}-\eqref{4.18}, we get
that
\begin{equation}\label{4.27}
\left\{\begin{aligned}   \xi'&=0,\\  g'&=-2\gamma g
\end{aligned}\right.
\end{equation} and
\begin{equation}\label{4.28}
\left\{\begin{aligned}
    &\phi_t-2\gamma\x\phi_{\x}+\gamma\phi-\nu\phi_{\x\x}-\psi\beta'\cos\alpha+\mu_0(\xi\sigma+g\sigma_{\x}-2\x\xi\sigma_{\x})\\
           &\ \ \ \ =((\beta')^2\sin\alpha\cos\alpha+\alpha'\gamma-\alpha'')\x+\theta_1,\\
   & \psi_y-2\gamma\x\psi_{\x}+\gamma\psi-\nu\psi_{\x\x}+\phi\beta'\cos\alpha+\mu_0(\xi h+gh_{\x}-2\x\xi
           h_{\x})\\&\ \ \ \ =-((\beta'\sin\alpha)'+\alpha'\beta'\cos\alpha+\alpha'\beta'\cos\alpha-\gamma'\beta'\sin\alpha)\x+\theta_2,\\
    &\sigma_t-2\gamma\x\sigma_{\x}-\gamma\sigma-\eta\sigma_{\x\x}-h\beta'\cos\alpha
           +(\xi\phi-g\phi_{\x}+2\x\xi\phi_{\x})\\&\ \ \ \ =3\alpha'\x\xi-\alpha'g,\\
    &\h_t-2\gamma\x h_{\x}-\gamma h-\eta
    h_{\x\x}+\sigma\beta'\cos\alpha+(\xi\psi-g\psi_{\x}+2\x\xi\psi_{\x})\\&\ \ \ \ =3\x\beta'\xi\sin\alpha-g\beta'\sin\alpha
    ,
\end{aligned}\right.
\end{equation}
where $\theta_1$ and $\theta_2$ are any functions of $t$.

Set
\begin{equation}\label{4.29}
 \varphi=\int\beta'\cos\alpha\md t\ \ \ \ \ \textrm{and} \ \ \ \mu=\exp(4\int\gamma\md
 t).
\end{equation} We have $g=c/\sqrt{\mu}$ by \eqref{4.27}, where $c$ is a constant.

Change variables by
\begin{equation}\label{4.30}
\begin{pmatrix}
  \phi\\ \psi\\ \sigma\\ h
\end{pmatrix}=\begin{pmatrix}
  \mu^{-\frac{1}{4}}\cos\varphi & \mu^{-\frac{1}{4}}\sin\varphi & 0 & 0\\
  -\mu^{-\frac{1}{4}}\sin\varphi & \mu^{-\frac{1}{4}}\cos\varphi & 0 & 0\\
  0 & 0 & \mu^{\frac{1}{4}}\cos\varphi & \mu^{\frac{1}{4}}\sin\varphi\\
  0 & 0 & -\mu^{\frac{1}{4}}\sin\varphi & \mu^{\frac{1}{4}}\cos\varphi\\
\end{pmatrix}\begin{pmatrix}
  \hat{\phi}\\ \hat{\psi}\\ \hat{\sigma}\\ \hat{h}
\end{pmatrix}.
\end{equation}
Then \eqref{4.28} becomes
\begin{equation}\label{4.31}
\left\{\begin{aligned}
   \hat{\phi}_t-\frac{\mu'}{2\mu}\x\hat{\phi}_{\x}-\nu\hat{\phi}_{\x\x}+\mu_0\sqrt{\mu}(\xi\hat{\sigma}+\frac{c}{\sqrt{\mu}}\hat{\sigma}_{\x}
      -2\xi\x\hat{\sigma}_{\x})&=q_1\x+\beta_1,\\
   \hat{\sigma}_t-\frac{\mu'}{2\mu}\x\hat{\sigma}_{\x}-\eta\hat{\sigma}_{\x\x}+\frac{\xi}{\sqrt{\mu}}\hat{\phi}
      -\frac{c}{\mu}\hat{\phi}_{\x}+2\frac{\xi}{\sqrt{\mu}}\x\hat{\phi}_{\x}&=q_2\x+\beta_2,\\
   \hat{\psi}_t-\frac{\mu'}{2\mu}\x\hat{\psi}_{\x}-\nu\hat{\psi}_{\x\x}+\mu_0\sqrt{\mu}(\xi\hat{h}+\frac{c}{\sqrt{\mu}}\hat{h}_{\x}
      -2\x\xi\hat{h}_{\x})&=q_3\x+\beta_3,\\
   \hat
   h_t-\frac{\mu'}{2\mu}\x\hat{h}_{\x}-\eta\hat{h}_{\x\x}+\frac{\xi}{\sqrt{\mu}}\hat{\psi}-\frac{c}{\mu}\hat{\psi}_{\x}
      +\frac{2\xi}{\sqrt{\mu}}\x\hat{\psi}_{\x}&=q_4\x+\beta_4,
\end{aligned}\right.
\end{equation}
where
\begin{equation}\label{4.32}
\left\{\begin{aligned}
    q_1&=\mu^{\frac{1}{4}}(((\beta')^2\sin\alpha\cos\alpha+\alpha'\gamma-\alpha'')\cos\varphi\
       -((\beta'\sin\alpha)'+(\alpha'-\gamma')\beta'\cos\alpha)\sin\varphi),\\
    q_2&=\mu^{\frac{1}{4}}(-((\beta')^2\sin\alpha\cos\alpha+\alpha'\gamma-\alpha'')\sin\varphi
       -((\beta'\sin\alpha)'+(\alpha'-\gamma')\beta'\cos\alpha)\cos\varphi),\\
    q_3&=\mu^{-\frac{1}{4}}(3\alpha'\xi\cos\varphi+3\beta'\xi\sin\alpha\sin\varphi),\\
    q_4&=\mu^{-\frac{1}{4}}(-3\alpha'\xi\sin\varphi+3\beta'\xi\sin\alpha\cos\varphi)
\end{aligned}\right.
\end{equation}
and
\begin{equation}\label{4.33}
\left\{\begin{aligned}
    \beta_1&=\mu^{\frac{1}{4}}(\theta_1\cos\varphi-\theta_2\sin\varphi),\\
    \beta_2&=\mu^{\frac{1}{4}}(-\theta_1\sin\varphi-\theta_2\cos\varphi),\\
    \beta_3&=\mu^{-\frac{1}{4}}(-\alpha'g\cos\varphi-\beta'\xi\sin\alpha\sin\varphi),\\
    \beta_4&=\mu^{-\frac{1}{4}}(\alpha'g\sin\varphi-\beta'\xi\sin\alpha\cos\varphi).
\end{aligned}\right.
\end{equation}

Let $n$ be a positive integer. We write
\begin{equation}\label{4.34}
\hat{\phi}=\sum_{m=0}^n a_m(t)\x^m,\ \ \ \ \hat{\sigma}=\sum_{m=0}^n
b_m(t)\x^m,\ \ \ \ \hat{\psi}=\sum_{m=0}^nc_m(t)\x^m,\ \ \ \
\hat{h}=\sum_{m=0}^n d_m(t)\x^m.
\end{equation} Substituting above expressions into \eqref{4.31}, we
obtain that
\begin{equation}\label{4.35}
\left\{\begin{aligned}
a_{m,t}-(m+1)\frac{\mu'}{\mu}a_m-\xi\mu_0\sqrt{\mu}(2m+1)b_m &=A_m(t),\\
b_{m,t}-(m+1)\frac{\mu'}{2\mu}b_m+(2m+3)\frac{\xi}{\sqrt{\mu}}a_m
&=B_m(t),\end{aligned}\right.
\end{equation}
where
\begin{equation}\label{4.36}
\left\{\begin{aligned}
   A_m&=-c\mu_0(m+1)b_{m+1}+(m+1)(m+2)\nu a_{m+2}+\delta_{m,1}q_1+\delta_{m,0}\beta_1,\\
   B_m&=\frac{c}{\mu}(m+1)a_{m+1}+(m+1)(m+2)\eta
   b_{m+2}+\delta_{m,1}q_2+\delta_{m,0}\beta_2\end{aligned}\right.
\end{equation} and
\begin{equation}\label{4.37}
\left\{\begin{aligned}
c_{m,t}-(m+1)\frac{\mu'}{\mu}c_m-\xi\mu_0\sqrt{\mu}(2m+1)d_m &=C_m(t),\\
d_{m,t}-(m+1)\frac{\mu'}{2\mu}d_m+(2m+3)\frac{\xi}{\sqrt{\mu}}c_m
&=D_m(t),\end{aligned}\right.
\end{equation}
where
\begin{equation}\label{4.38}
\left\{\begin{aligned}
   C_m&=-c\mu_0(m+1)d_{m+1}+(m+1)(m+2)\nu c_{m+2}+\delta_{m,1}q_3+\delta_{m,0}\beta_3,\\
   D_m&=\frac{c}{\mu}(m+1)c_{m+1}+(m+1)(m+2)\eta
   d_{m+2}+\delta_{m,1}q_4+\delta_{m,0}\beta_4.\end{aligned}\right.
\end{equation}
As we did before, let
\begin{equation}\label{4.39}
\mu=l_1\exp(\int(2k+\frac{4}{-1+2l_2e^{-kt}})\md t)
\end{equation} for any constants $k$, $l_1$ and $l_2$. Under this
condition, the equations \eqref{4.35} and \eqref{4.37} may be solved
as second order constant-coefficient ordinary differential equations
via certain changing of variables. We only give the results but omit
the procedures.

Write $\lambda_m=\frac{1}{4}k^2-(2m+1)(2m+3)\xi^2\mu_0$ and define
\begin{equation}\label{4.40}
 \zeta_{m,1}=\left\{ \begin{array}{ll}
                            \mu\exp(\sqrt{\lambda_m}\ t) & \textrm{if }\lambda_m>0,\\
                            \mu & \textrm{if }\lambda_m=0,\\
                            \mu\cos(i\sqrt{\lambda_m}\ t)& \textrm{if
                              }\lambda_m<0,
                       \end{array}\right.
\end{equation}
\begin{equation}\label{4.41}
 \zeta_{m,2}=\left\{ \begin{array}{ll}
                            \mu\exp(-\sqrt{\lambda_m}\ t) & \textrm{if }\lambda_m>0,\\
                            t\mu& \textrm{if }\lambda_m=0,\\
                            \mu\sin(i\sqrt{\lambda_m}\ t)& \textrm{if
                              }\lambda_m<0.
                       \end{array}\right.
\end{equation}
In addition, we denote
\begin{equation}\label{4.42}
\left\{\begin{aligned}
   P_m&=A_m'-A_m-\frac{\mu'}{\mu}(m+1)A_m+\xi\mu_0(2m+1)\sqrt{\mu}B_m,\\
   Q_m&=C_m'-C_m-\frac{\mu'}{\mu}(m+1)C_m+\xi\mu_0(2m+1)\sqrt{\mu}D_m.
\end{aligned}
\right.
\end{equation} Then we get the following recursive formulae:
\begin{equation}\label{4.43}
   a_m=\sum_{j=1}^2\varsigma_{m,j}(q_{m,j}+\int_0^t\frac{W_j[\varsigma_{m,1}(s),\varsigma_{m,2}(s)]}{W[\varsigma_{m,1}(s),\varsigma_{m,2}(s)]}P_m(s)\md
s),
\end{equation}
\begin{eqnarray}\label{4.44}
 &&b_m=\frac{1}{(2m+1)\xi\mu_0\sqrt{\mu}}\sum_{j=1}^2\Big(
         \varsigma_{m,j}'(q_{m,j}+\int_0^t\frac{W_j[\varsigma_{m,1}(s),\varsigma_{m,2}(s)]}{W[\varsigma_{m,1}(s),\varsigma_{m,2}(s)]}P_m(s)\md
         s)\nonumber\\
    &+&\varsigma_{m,j}(\frac{W_j[\varsigma_{m,1}(t),\varsigma_{m,2}(t)]}{W[\varsigma_{m,1}(t),\varsigma_{m,2}(t)]}P_m(t)-
         (m+1)\frac{\mu'}{2\mu}(q_{m,j}+\int_0^t\frac{W_j[\varsigma_{m,1}(s),\varsigma_{m,2}(s)]}{W[\varsigma_{m,1}(s),\varsigma_{m,2}(s)]}P_m(s)\md
         s))\Big),\nonumber\\
\end{eqnarray}
\begin{equation}\label{4.45}
   c_m=\sum_{j=1}^2\varsigma_{m,j}(q_{m,j}+\int_0^t\frac{W_j[\varsigma_{m,1}(s),\varsigma_{m,2}(s)]}{W[\varsigma_{m,1}(s),\varsigma_{m,2}(s)]}Q_m(s)\md
s),
\end{equation} and
\begin{eqnarray}\label{4.46}
 &&d_m=\frac{1}{(2m+1)\xi\mu_0\sqrt{\mu}}\sum_{j=1}^2\Big(
         \varsigma_{m,j}'(q_{m,j}+\int_0^t\frac{W_j[\varsigma_{m,1}(s),\varsigma_{m,2}(s)]}{W[\varsigma_{m,1}(s),\varsigma_{m,2}(s)]}Q_m(s)\md
         s)\nonumber\\
    &+&\varsigma_{m,j}(\frac{W_j[\varsigma_{m,1}(t),\varsigma_{m,2}(t)]}{W[\varsigma_{m,1}(t),\varsigma_{m,2}(t)]}Q_m(t)-
         (m+1)\frac{\mu'}{2\mu}(q_{m,j}+\int_0^t\frac{W_j[\varsigma_{m,1}(s),\varsigma_{m,2}(s)]}{W[\varsigma_{m,1}(s),\varsigma_{m,2}(s)]}Q_m(s)\md
         s))\Big),\nonumber\\
\end{eqnarray}
 where $q_{m,1}$ and $q_{m,2}$ are real constants.

\begin{thm}
Let $\alpha$ and $\beta$ be arbitrary functions of $t$, and let $c$
and $\xi$ be arbitrary real constants. The functions $\varphi$,
$\mu$, $\hat{\phi}$, $\hat\sigma$, $\hat\psi$ and $\hat h$ are given
in \eqref{4.29}, \eqref{4.39}, \eqref{4.34} and
\eqref{4.43}-\eqref{4.46}, respectively. Then we have the following
solutions of the MHD equations \eqref{M1}-\eqref{M4}:
 \begin{equation}\label{4.47}
 \left\{
 \begin{aligned}
   u=&-\x\frac{2\mu'}{\mu}\cos\alpha-\y(\alpha'\cos\alpha+\frac{\mu'}{4\mu}\sin\alpha)-\z\beta'\sin\alpha\cos\alpha
      -\mu^{-\frac{1}{4}}(\hat\phi\cos\varphi-\hat\psi\sin\varphi)\sin\alpha,\\
   v=&-\x\frac{2\mu'}{\mu}\sin\alpha\cos\beta+\y(-\alpha'\sin\alpha+\frac{\mu'}{4\mu}\cos\alpha)\cos\beta
          +\z(-\beta'\sin\alpha^2\cos\beta+\frac{\mu'}{4\mu}\sin\beta)\\
      &+\mu^{-\frac{1}{4}}((\cos\alpha\cos\beta\cos\varphi+\sin\beta\sin\varphi)\hat\phi
      +(\cos\alpha\cos\beta\sin\varphi-\sin\beta\cos\varphi)\hat\psi),\\
   w=&-\x\frac{2\mu'}{\mu}\sin\alpha\sin\beta+\y(-\alpha'\sin\alpha+\frac{\mu'}{4\mu}\cos\alpha)\sin\beta
          +\z(-\beta'\sin\alpha^2\sin\beta+\frac{\mu'}{4\mu}\cos\beta)\\
      &+\mu^{-\frac{1}{4}}((\cos\alpha\sin\beta\cos\varphi-\cos\beta\sin\varphi)\hat\phi+(\cos\alpha\sin\beta\sin\varphi+\cos\beta\cos\varphi)\hat\psi),
 \end{aligned}\right.
 \end{equation}
\begin{equation}\label{4.48}
\left\{ \begin{aligned}
  H^1&=\frac{c}{\sqrt\mu}\cos\alpha-2\x\xi\cos\alpha-\y\xi\sin\alpha-\mu^{\frac{1}{4}}\sin\alpha(\hat\sigma\cos\varphi+\hat h\sin\varphi),\\
  H^2&=\frac{c}{\sqrt\mu}\sin\alpha\cos\beta-2\x\xi\sin\alpha\cos\beta+\y\xi\cos\alpha\cos\beta-\z\xi\sin\beta\\
  &+\mu^{-\frac{1}{4}}((\cos\alpha\cos\beta\cos\varphi+\sin\beta\sin\varphi)\hat\sigma+(\cos\alpha\cos\beta\sin\varphi-\sin\beta\cos\varphi)\hat h),\\
  H^3&=\frac{c}{\sqrt\mu}\sin\alpha\sin\beta-2\x\xi\sin\alpha\sin\beta+\y\xi\cos\alpha\cos\beta+\z\xi\cos\beta\\
  &+\mu^{-\frac{1}{4}}((\cos\alpha\sin\beta\cos\varphi-\cos\beta\sin\varphi)\hat\sigma+(\cos\alpha\sin\beta\sin\varphi+\cos\beta\cos\varphi)\hat
  h),
\end{aligned}\right.
\end{equation}
\begin{eqnarray}\label{4.49}
p&=&-\rho(\frac{-2\mu^2\mu''-(\mu')^3+\mu\mu'\mu''+2\mu(\mu')^2}{4\mu^3}+(\alpha')^2
+(\beta'\sin\alpha)^2)\frac{\x^2}{2}+(\frac{(\mu')^2}{16\mu^2}-(\alpha')^2)\frac{\y^2}{2}\nonumber\\
&&+(\frac{(\mu')^2}{16\mu^2}-(\beta'\sin\alpha)^2)\frac{\z^2}{2}
-(\alpha''+\frac{\mu'}{2\mu}\alpha'-\beta'\sin\alpha\beta'\cos\alpha)\x\y\nonumber\\
&&-((\beta'\sin\alpha)'+\alpha'\beta'\cos\alpha+\frac{\mu'}{2\mu}\beta'\sin\alpha)\x\z-\y\z\alpha'\beta'\sin\alpha-2\alpha'\int\phi\md\x\nonumber\\
&&-2\beta'\sin\alpha\int\psi\md\x-\mu_0(\frac{1}{2}(\sigma^2+h^2)+\xi\sigma\y+\xi,
h\z)
\end{eqnarray}where $\phi$, $\psi$, $\sigma$ and $h$ are given by
\eqref{4.30}.
\end{thm}

\vskip 1.0cm

Set
\begin{equation}\label{4.50}
\left\{ \begin{aligned}
  \U&=a_{11}\x+a_{12}\y+a_{13}\z-6\nu\x^{-1},\\
  \V&=a_{21}\x+a_{22}\y+a_{23}\z-6\nu\y\x^{-2},\\
  \W&=a_{31}\x+a_{32}\y+a_{33}\z,
\end{aligned}\right.\ \ \ \ \ \ \ \
\left\{ \begin{aligned}
  \h^1&=b_{11}\x+b_{12}\y+b_{13}\z,\\
  \h^2&=b_{21}\x+b_{22}\y+b_{23}\z,\\
  \h^3&=b_{31}\x+b_{32}\y+b_{33}\z.
\end{aligned}\right.
\end{equation}
Substituting \eqref{4.50} into \eqref{4.18}, one gets that
\begin{equation}\label{4.51}
a_{11}=a_{22}=\gamma,\ \ a_{21}=-a_{12}=\alpha',\ \
a_{31}=-a_{13}=\beta'\sin\alpha,\ \ a_{23}=a_{32}=-\beta'\cos\alpha
\end{equation} and
\begin{equation}\label{4.52}\left\{\begin{aligned}
2\alpha''+4\alpha'\gamma&=\mu_0(b_{31}(b_{32}-b_{23})+b_{33}(b_{21}-b_{12})+b_{32}(b_{13}-b_{31})),\\
b_{22}(b_{13}-b_{31})&=-b_{23}(b_{21}-b_{12})-b_{21}(b_{32}-b_{23}),\\
b_{11}(b_{32}-b_{23})&=b_{12}(b_{32}-b_{23})-b_{13}(b_{21}-b_{12}).
\end{aligned}\right.\end{equation}
By \eqref{4.16}, we have that
\begin{eqnarray}\label{4.53}
&&b_{11}'\x+b_{12}'\y+b_{13}'\z+b_{11}(\alpha'\y+\z\beta'\sin\alpha)+b_{12}(\gamma\y-6\nu\y\x^{-2})\nonumber\\
&-&2(\x\beta'\sin\alpha+\y\beta'\cos\alpha+\gamma\z)b_{13}
-(\gamma+6\nu\x^{-2})(b_{11}\x+b_{12}\y+b_{13}\z)\nonumber\\
&+&b_{11}(\gamma\x-\alpha'\y-\z\beta'\sin\alpha-6\nu\x^{-1})=0,
\end{eqnarray}
\begin{eqnarray}\label{4.54}
&&b_{21}'\x+b_{22}'\y+b_{23}'\z+b_{21}(\gamma\x-6\nu\x^{-1})+b_{22}(-\alpha'\x+\z\beta'\cos\alpha)\nonumber\\
&-&2(\x\beta'\sin\alpha+\y\beta'\cos\alpha+\gamma\z)b_{23}+(\gamma+6\nu\x^{-2})(b_{12}\x+b_{22}\y+b_{23}\z)\nonumber\\
&+&b_{22}(\alpha'\x+\gamma\y-\z\beta'\cos\alpha-6\nu\y\x^{-2})=0
\end{eqnarray} and
\begin{eqnarray}\label{4.55}
&&b_{31}'\x+b_{32}'\y+b_{33}'\z+b_{31}(\gamma\x-6\nu\x^{-1})+b_{32}(\gamma\y-6\nu\y\x^{-2})\nonumber\\
&-&b_{33}(\x\beta'\sin\alpha+\y\beta'\cos\alpha)+2(b_{21}\x+b_{22}\y)\beta'\cos\alpha\nonumber\\
&+&2\gamma(b_{31}\x+b_{32}\y+b_{33}\z)-b_{33}(\x\beta'\sin\alpha+\y\beta'\cos\alpha+2\gamma\z)=0.
\end{eqnarray}
One shows that $\widetilde{\textbf{H}}\neq0$ if and only if $\beta$
is a constant. In this case,
\begin{equation}\label{4.56}
b_{11}=b_{12}=b_{13}=b_{23}=b_{31}=b_{32}=0
\end{equation} and
\begin{equation}\label{4.57}
   b_{22}'=b_{21}'=0.
\end{equation} Hence, one gets
\begin{equation}\label{4.58}
\gamma=-\frac{\alpha''}{2\alpha'}-\frac{\mu_0
b_{22}b_{21}}{4\alpha'}.
\end{equation}

\begin{thm}
  For any function $\alpha(t)$ and constants $a$,
  $b$, we have a solution of the MHD equations \eqref{M1}-\eqref{M4}:
  \begin{equation}\label{4.59}
  \left\{
  \begin{aligned}
    u=&-(\frac{\alpha''}{2\alpha'}+\frac{\mu_0
         ab}{4\alpha'})x-\alpha'y-\frac{6\nu\cos\alpha}{x\cos\alpha+y\sin\alpha}
         +6\nu\sin\alpha\frac{-x\sin\alpha+y\cos\alpha}{(x\cos\alpha+y\sin\alpha)^2},\\
    v=&\alpha'x-(\frac{\alpha''}{2\alpha'}+\frac{\mu_0
         ab}{4\alpha'})y-\frac{6\nu\sin\alpha}{x\cos\alpha+y\sin\alpha}-
         6\nu\cos\alpha\frac{-x\sin\alpha+y\cos\alpha}{(x\cos\alpha+y\sin\alpha)^2},\\
    w=&2(\frac{\alpha''}{2\alpha'}+\frac{\mu_0
         ab}{4\alpha'})z,
  \end{aligned}\right.
  \end{equation}
  \begin{equation}\label{4.60} \left\{
  \begin{aligned}
  H^1&=-[(a\cos\alpha-b\sin\alpha)x+(-a\sin\alpha+b\cos\alpha)y]\sin\alpha,\\
  H^2&=[(a\cos\alpha-b\sin\alpha)x+(a\sin\alpha+b\cos\alpha)y]\cos\alpha,\\
  H^3&=-bz
  \end{aligned}\right.\end{equation} and
  \begin{eqnarray}\label{4.61}
  p&=&-\rho\Big(((\frac{\alpha''}{2\alpha'}+\frac{\mu_0
         ab}{4\alpha'})'+(\frac{\alpha''}{2\alpha'}+\frac{\mu_0
         ab}{4\alpha'})^2-(\alpha')^2-\mu_0a^2)\frac{(x\cos\alpha+y\sin\alpha)^2}{2}\nonumber\\
         &&+((\frac{\alpha''}{2\alpha'}+\frac{\mu_0
         ab}{4\alpha'})'+(\frac{\alpha''}{2\alpha'}+\frac{\mu_0
         ab}{4\alpha'})^2-(\alpha')^2)\frac{(-x\sin\alpha+y\cos\alpha)^2}{2}\nonumber\\
      &&-((\frac{\alpha''}{2\alpha'}+\frac{\mu_0
         ab}{4\alpha'})'+2(\frac{\alpha''}{2\alpha'}+\frac{\mu_0
         ab}{4\alpha'})^2)\frac{z^2}{2}\nonumber\\
      &&+(\alpha''+2\alpha'(\frac{\alpha''}{2\alpha'}+\frac{\mu_0
         ab}{4\alpha'}))(x\cos\alpha+y\sin\alpha)(-x\sin\alpha+y\cos\alpha)\nonumber\\
         &&-12\nu\alpha'\frac{-x\sin\alpha+y\cos\alpha}
         {x\cos\alpha+y\sin\alpha}+\frac{12\nu^2}
         {(x\cos\alpha+y\sin\alpha)^2}\Big).
  \end{eqnarray}
\end{thm}

\vskip 1.0cm

Let $\alpha_1$, $\beta_1$ and $\gamma$ be functions of $t$ and $a$,
$b$ be real constants. Set
\begin{equation}\label{4.62}
\xi_r=\left\{\begin{array}{ll}
      e^{\alpha_1\y+\beta_1\z}-ae^{-(\alpha_1\y+\beta_1\z)} & \textrm{if }r=0,\\
      \sin(\alpha_1\y+\beta_1\z) & \textrm{if }r=1;
\end{array}\right.
\end{equation}
\begin{equation}\label{4.63}
\zeta_r=\left\{\begin{array}{ll}
      e^{\alpha_1\y+\beta_1\z}+ae^{-(\alpha_1\y+\beta_1\z)} & \textrm{if }r=0,\\
      \cos(\alpha_1\y+\beta_1\z) & \textrm{if }r=1;
\end{array}\right.
\end{equation}
\begin{equation}\label{4.64}
\phi_s=\left\{\begin{array}{ll}
      e^{\gamma\x}-be^{-\gamma\x} & \textrm{if }s=0,\\
      \sin(\gamma\x) & \textrm{if }s=1
\end{array}\right.\ \ \textrm{and }\psi_s=\left\{\begin{array}{ll}
      e^{\gamma\x}+be^{-\gamma\x} & \textrm{if }s=0,\\
      \cos(\gamma\x) & \textrm{if }s=1.
\end{array}\right.
\end{equation}

We assume that
\begin{equation}\label{4.65}
\begin{pmatrix}
  \U\\\V\\\W
\end{pmatrix}=A \begin{pmatrix}
  \x\\\y\\\z
\end{pmatrix}+ \begin{pmatrix}
  -(\alpha_1\sigma+\beta_1\tau)\zeta_r\phi_s\\
  \sigma\gamma\xi_r\psi_s\\
  \tau\gamma\xi_r\psi_s
\end{pmatrix},\end{equation}
\begin{equation}\label{4.66}
\begin{pmatrix}
  \h^1\\\h^2\\\h^3
\end{pmatrix}=B \begin{pmatrix}
  \x\\\y\\\z
\end{pmatrix}+ \begin{pmatrix}
  -(\alpha_1\sigma_1+\beta_1\tau_1)\zeta_r\phi_s\\
  \sigma_1\gamma\xi_r\psi_s\\
  \tau_1\gamma\xi_r\psi_s
\end{pmatrix},
\end{equation} where $A=(a_{i,j})_{3\times3}$ and $B=(b_{i,j})_{3\times3}$ are $3\times 3$ matrices whose
entries are functions of $t$; the functions $\sigma$, $\tau$,
$\sigma_1$ and $\tau_1$ are also functions of $t$; the numbers $r$
and $s$ may be 0 or 1.

Now
\begin{eqnarray}\label{4.67}
R_1&=&\ \ (a_{11}'+a_{11}^2+a_{12}(a_{21}-\alpha')+a_{13}(a_{31}-\beta'\sin\alpha)-a_{21}\alpha'-a_{31}\beta'\sin\alpha)\x\nonumber\\
   &&+(a_{12}'+a_{11}(a_{12}+\alpha')+a_{12}a_{22}+a_{13}(a_{32}-\beta'\cos\alpha)-a_{22}\alpha'-a_{32}\beta'\sin\alpha)\y\nonumber\\
   &&+(a_{13}'+a_{11}(a_{13}+\beta'\sin\alpha)+a_{12}(a_{23}+\beta'\cos\alpha)+a_{13}a_{33}-a_{23}\alpha'-a_{33}\beta'\sin\alpha)\z\nonumber\\
   &&+(\delta_{r,1}+4a\delta_{r,0})(\alpha_1\sigma+\beta_1\tau)^2\gamma\phi_s\psi_s\nonumber\\
   &&-(-1)^r(\alpha_1\sigma+\beta_1\tau)[\alpha_1'\y+\beta_1'\z+\alpha_1((a_{21}-\alpha')\x+a_{22}\y+(a_{23}+\beta'\cos\alpha)\z)\nonumber\\
   &&+\beta_1((a_{31}-\beta'\sin\alpha)\x+(a_{32}-\beta'\cos\alpha)\y+a_{33}\z)]\xi_r\phi_s\nonumber\\
   &&+\gamma[(a_{12}-\alpha')\sigma+(a_{13}-\beta'\sin\alpha)\tau]\xi_r\psi_s\nonumber\\
   &&-[(\alpha_1\sigma+\beta_1\tau)'+a_{11}(\alpha_1\sigma+\beta_1\tau)+\nu(\alpha_1\sigma+\beta_1\tau)
      ((-1)^s\gamma^2+(-1)^r(\alpha_1^2+\beta_1^2))]\zeta_r\phi_s\nonumber\\
   &&-(\alpha_1\sigma+\beta_1\tau)[\gamma'\x+\gamma(a_{11}\x+(a_{12}+\alpha')\y+(a_{13}+\beta'\sin\alpha)\z)]\zeta_r\psi_s,
\end{eqnarray}
\begin{eqnarray}\label{4.68}
R_2&=&\ \ (a_{21}'+a_{21}a_{11}+a_{22}(a_{21}-\alpha')+a_{23}(a_{31}-\beta'\sin\alpha)+a_{11}\alpha'-a_{31}\beta'\cos\alpha)\x\nonumber\\
   &&+(a_{22}'+a_{21}(a_{12}+\alpha')+a_{22}^2+a_{23}(a_{32}-\beta'\cos\alpha)+a_{12}\alpha'-a_{32}\beta'\cos\alpha)\y\nonumber\\
   &&+(a_{23}'+a_{21}(a_{13}+\beta'\sin\alpha)+a_{22}(a_{23}+\beta'\cos\alpha)+a_{23}a_{33}+a_{13}\alpha'-a_{33}\beta'\cos\alpha)\z\nonumber\\
   &&+(\delta_{r,1}+4a\delta_{r,0})\sigma\gamma^2(\alpha_1\sigma+\beta_1\tau)\xi_r\zeta_r\nonumber\\
   &&+(-1)^s\sigma\gamma[\gamma'\x+\gamma(a_{11}\x+(a_{12}+\alpha')\y+(a_{13}+\beta'\sin\alpha)\z)]\xi_r\phi_s\nonumber\\
   &&+[(\sigma\gamma)'+a_{22}\sigma\gamma+(a_{23}-\beta'\cos\alpha)\tau\gamma
      -\nu\sigma\gamma((-1)^s\gamma^2+(-1)^r(\alpha_1^2+\beta_1^2))]\xi_r\psi_s\nonumber\\
   &&-(a_{21}+\alpha')(\alpha_1\sigma+\beta_1\tau)\zeta_r\phi_s\nonumber\\&&+\sigma\gamma[(\alpha_1'\y+\beta_1'\z)+
       \alpha_1((a_{21}-\alpha')\x+a_{22}\y+(a_{23}+\beta'\cos\alpha)\z)\nonumber\\
   &&+\beta_1((a_{31}-\beta'\sin\alpha)\x+(a_{32}-\beta'\cos\alpha)\y+a_{33}\z)]\zeta_r\psi_s,
\end{eqnarray}
\begin{eqnarray}\label{4.69}
R_3&=&\ \ (a_{31}'+a_{31}a_{11}+a_{32}(a_{21}-\alpha')+a_{33}(a_{31}-\beta'\sin\alpha)+a_{11}+a_{21}\beta'\cos\alpha)\x\nonumber\\
   &&+(a_{32}'+a_{31}(a_{12}+\alpha')+a_{32}a_{22}+a_{33}(a_{32}-\beta'\cos\alpha)+a_{12}+a_{22}\beta'\cos\alpha)\y\nonumber\\
   &&+(a_{33}'+a_{31}(a_{13}+\beta'\sin\alpha)+a_{32}(a_{23}+\beta'\cos\alpha)+a_{33}^2+a_{13}\beta'\sin\alpha+a_{23}\beta'\cos\alpha)\z\nonumber\\
   &&+(\delta_{r,1}+4a\delta_{r,0})\tau\gamma^2(\alpha_1\sigma+\beta_1\tau)\xi_r\zeta_r\nonumber\\
   &&+(-1)^s\tau\gamma[\gamma'\x+\gamma(a_{11}\x+(a_{12}+\alpha')\y+(a_{13}+\beta'\sin\alpha)\z)]\xi_r\phi_s\nonumber\\
   &&+[(\tau\gamma)'+(a_{32}+\beta'\cos\alpha)\sigma\gamma+a_{33}\tau\gamma
      -\nu\tau\gamma((-1)^s\gamma^2+(-1)^r(\alpha_1^2+\beta_1^2))]\xi_r\psi_s\nonumber\\
   &&-(a_{31}+\beta'\sin\alpha)(\alpha_1\sigma+\beta_1\tau)\zeta_r\phi_s\nonumber\\&&+\tau\gamma[(\alpha_1'\y+\beta_1'\z)+
       \alpha_1((a_{21}-\alpha')\x+a_{22}\y+(a_{23}+\beta'\cos\alpha)\z)\nonumber\\
   &&+\beta_1((a_{31}-\beta'\sin\alpha)\x+(a_{32}-\beta'\cos\alpha)\y+a_{33}\z)]\zeta_r\psi_s,
\end{eqnarray}
\begin{eqnarray}\label{4.70}
G_1&=&(b_{21}-b_{12})(b_{21}\x+b_{22}\y+b_{23}\z)+(b_{31}-b_{13})(b_{31}\x+b_{32}\y+b_{33}\z)\nonumber\\
   &&+\gamma[(-1)^s\gamma^2(\sigma_1^2+\tau_1^2)+(-1)^r(\alpha_1\sigma_1+\beta_1\tau_1)^2]\xi_r^2\phi_s\psi_s\nonumber\\
   &&+[(b_{21}\x+b_{22}\y+b_{23}\z)((-1)^s\sigma_1\gamma^2+(-1)^r\alpha_1(\alpha_1\sigma_1+\beta_1\tau_1))\nonumber\\
   &&+(b_{31}\x+b_{32}\y+b_{33}\z)((-1)^s\tau_1\gamma^2+(-1)^r\beta_1(\alpha_1\sigma_1+\beta_1\tau_1))]\xi_r\phi_s\nonumber\\
   &&+\gamma[\sigma_1(b_{21}-b_{12})+\tau_1(b_{31}-b_{13})]\xi_r\psi_s,
\end{eqnarray}
\begin{eqnarray}\label{4.71}
G_2&=&(b_{32}-b_{23})(b_{31}\x+b_{32}\y+b_{33}\z)-(b_{21}-b_{12})(b_{11}\x+b_{12}\y+b_{13}\z)\nonumber\\
   &&+(\alpha_1\sigma_1+\beta_1\tau_1)((-1)^s\sigma_1\gamma^2+(-1)^r\alpha_1(\alpha_1\sigma_1+\beta_1\tau_1))\xi_r\zeta_r\phi_s^2\nonumber\\
   &&+\gamma^2\tau_1(\alpha_1\tau_1-\beta_1\sigma_1)\xi_r\zeta_r\psi_s^2\nonumber\\
   &&-(b_{11}\x+b_{12}\y+b_{13}\z)((-1)^s\sigma_1\gamma^2+(-1)^r\alpha_1(\alpha_1\sigma_1+\beta_1\tau_1))\xi_r\phi_s\nonumber\\
   &&+(b_{32}-b_{23})\tau_1\gamma\xi_r\psi_s-(b_{21}-b_{12})(\alpha_1\sigma_1+\beta_1\tau_1)\zeta_r\phi_s\nonumber\\
   &&+(b_{31}\x+b_{32}\y+b_{33}\z)(\alpha_1\tau_1-\beta_1\sigma_1)\gamma\zeta_r\psi_s,
\end{eqnarray}
\begin{eqnarray}\label{4.72}
G_3&=&(b_{13}-b_{31})(b_{11}\x+b_{12}\y+b_{13}\z)-(b_{32}-b_{23})(b_{21}\x+b_{22}\y+b_{23}\z)\nonumber\\
   &&+(\alpha_1\sigma_1+\beta_1\tau_1)((-1)^s\tau_1\gamma^2+(-1)^r\beta1(\alpha_1\sigma_1+\beta_1\tau_1))\xi_r\zeta_r\phi_s^2\nonumber\\
   &&-\gamma^2\sigma_1(\alpha_1\tau_1-\beta_1\sigma_1)\xi_r\zeta_r\psi_s^2\nonumber\\
   &&-(b_{11}\x+b_{12}\y+b_{13}\z)((-1)^s\sigma_1\gamma^2+(-1)^r\beta_1(\alpha_1\sigma_1+\beta_1\tau_1))\xi_r\phi_s\nonumber\\
   &&-(b_{32}-b_{23})\sigma_1\gamma\xi_r\psi_s-(b_{13}-b_{31})(\alpha_1\sigma_1+\beta_1\tau_1)\zeta_r\phi_s\nonumber\\
   &&-(b_{21}\x+b_{22}\y+b_{23}\z)(\alpha_1\tau_1-\beta_1\sigma_1)\gamma\zeta_r\psi_s.
\end{eqnarray}
We assume that $\alpha_1\sigma_1+\beta_1\tau_1\neq0$ and
$\alpha_1\sigma+\beta_1\tau\neq0$. Then the coefficients of
$\zeta_r^2\phi_r^2$ in $\partial_{\y}\Phi_1$ and
$\partial_{\x}\Phi_2$ yield
\begin{equation}\label{4.73}
\alpha_1\tau_1=\beta_1\sigma_1.
\end{equation}

For simplicity, we consider the special case in which
\begin{equation}\label{4.74}
a_{21}=-a_{12}=\alpha',\ \ a_{31}=-a_{13}=\beta'\sin\alpha,\ \
a_{23}=a_{32}\ \ \textrm{and }\ B=B^t.
\end{equation}
Furthermore, the nonlinear terms in $\Phi_{1\y}$, $\Phi_{1\z}$,
$\Phi_{2\x}$, $\Phi_{2\z}$, $\Phi_{3\x}$ and $\Phi_{3\y}$ hint us
that it is convenient to suppose
\begin{equation}\label{4.75}
\gamma^2=(-1)^{r+s+1}(\alpha_1^2+\beta_1^2).
\end{equation}
Thus the coefficients of $\xi_r\psi_s$ in $\Phi_{2\y}$ and
$\Phi_{2\z}$ suggest that
\begin{equation}\label{4.76}
\alpha_1\tau=\beta_1\sigma.
\end{equation}
Hence we can write
\begin{equation}\label{4.77}
\sigma_1=g\sigma,\ \ \ \ \ \ \ \ \tau_1=g\tau.
\end{equation} As Xu did in \cite{X1}, we get that
\begin{equation}\label{4.78}
\left\{\begin{aligned}
           \alpha_1&=\varphi\alpha',\\
           \beta_1&=\varphi\beta'\sin\alpha
\end{aligned}\right.\ \ \textrm{and }\ \
\left\{\begin{aligned}
           \sigma&=\mu\alpha',\\
           \tau&=\mu\beta'\sin\alpha
\end{aligned}\right.
\end{equation} for some functions
$\varphi$ and $\mu$. Moreover,
\begin{equation}\label{4.79}
a_{22}=\frac{-a_{11}(\alpha')^2-\alpha'\alpha''+\beta'\sin\alpha\
(\beta'\sin\alpha)'+2\alpha'\beta'\sin\alpha\ \beta'\cos\alpha}
{(\alpha')^2+(\beta'\sin\alpha)^2},
\end{equation}
\begin{equation}\label{4.80}
a_{33}=\frac{-a_{11}(\beta'\sin\alpha)^2+\alpha'\alpha''-\beta'\sin\alpha\
(\beta'\sin\alpha)'-2\alpha'\beta'\sin\alpha\ \beta'\cos\alpha}
{(\alpha')^2+(\beta'\sin\alpha)^2}
\end{equation} and
\begin{equation}\label{4.81}
a_{23}=\frac{-a_{11}\alpha'\beta'\sin\alpha-\alpha''\beta'\sin\alpha-\alpha'\beta''\sin\alpha-2(\alpha')^2\beta'\cos\alpha+(\beta')^3\sin^2\alpha\
\cos\alpha} {(\alpha')^2+(\beta'\sin\alpha)^2}.
\end{equation}
The coefficients of $\zeta_r\psi_s$ and $\y$ in $\Psi_1=0$ show that
\begin{equation}\label{4.82}
b_{11}=b_{12}=b_{13}=0.
\end{equation} Substituting \eqref{4.74}-\eqref{4.82} into
\eqref{4.16}, one gets that
\begin{equation}\label{4.83}
a_{11}=-\frac{\gamma'}{\gamma}=\frac{(g(\alpha_1\sigma+\beta_1\tau))'}{g(\alpha_1\sigma+\beta_1\tau)},
\end{equation}
\begin{equation}\label{4.84}
b_{22}(a_{23}+a_{32})=b_{33}(a_{22}-a_{33}),
\end{equation}
\begin{equation}\label{4.85}
\left\{\begin{aligned}
   b_{22}'&=2b_{23}\beta'\cos\alpha,\\
   b_{23}'&=-2b_{22}\beta'\cos\alpha,
\end{aligned}\right.
\end{equation}
\begin{equation}\label{4.86}
\left\{\begin{aligned}
      b_{22}&=\frac{g}{\gamma^2}(\alpha_1\alpha_1'-\beta_1\beta_1'+\alpha_1^2a_{22}-\beta_1\beta_1^2a_{33}-2\alpha_1\beta_1\beta'\cos\alpha),\\
      b_{23}&=\frac{g}{\gamma^2}((\alpha_1\beta_1)'+\alpha_1\beta_1(a_{22}+a_{33})+(a_{23}+\beta'\cos\alpha)\tau^2\gamma g+(a_{23}
      -\beta'\cos\alpha)\sigma^2\gamma g)
\end{aligned}\right.
\end{equation}
and
\begin{equation}\label{4.87}
\left\{\begin{aligned}
   b_{22}&=\frac{-\sigma(g\sigma\gamma)'+\tau(g\tau\gamma)'+a_{22}\sigma^2\gamma g-a_{33}\tau^2\gamma g+2\sigma\tau\gamma g\beta'\cos\alpha}
   {\gamma(\sigma^2+\tau^2)},\\
   b_{23}&=\frac{-\tau(g\sigma\gamma)'+\sigma(g\tau\gamma)'-a_{11}\sigma\tau\gamma g+(a_{23}+\beta'\cos\alpha)\tau^2\gamma g+(a_{23}
      -\beta'\cos\alpha)\sigma^2\gamma g}
   {\gamma(\sigma^2+\tau^2)}.
\end{aligned}\right.
\end{equation}
One shows that the above systems \eqref{4.84}-\eqref{4.87} are
compatible with each other if and only if
\begin{equation}\label{4.88}
\alpha''\beta'\sin\alpha-\alpha'\beta''\sin\alpha-2(\alpha')^2\beta'\cos\alpha-(\beta')^3\sin^2\alpha\
\cos\alpha=0,
\end{equation} i.e.
\begin{equation}\label{4.89}
\beta=\pm\int\frac{\md \alpha}{\sin^2\alpha\
\sqrt{d-\sin^{-2}\alpha}}+d_0
\end{equation} for some constants $d_0$ and $d>1$. Moreover, the system
\eqref{4.86} says that
\begin{equation}\label{4.90}
\left\{\begin{aligned}
   b_{22}&=\frac{g}{Q}(\frac{2\varphi'}{\varphi}+\frac{Q'}{2Q})((\alpha')^2-(\beta'\sin\alpha)^2),\\
   b_{23}&=\frac{g}{Q}(\frac{4\varphi'}{\varphi}+\frac{Q'}{Q})\alpha'\beta'\sin\alpha,
\end{aligned}\right.
\end{equation} where
\begin{equation}\label{4.91}
Q=(\alpha')^2+(\beta'\sin\alpha)^2.
\end{equation} Substituting \eqref{4.90} into \eqref{4.85}, one
finds that $g(\frac{2\varphi'}{\varphi}+\frac{Q'}{2Q})$ is a
constant. We write
\begin{equation}\label{4.92}
g(\frac{2\varphi'}{\varphi}+\frac{Q'}{2Q})=2c.
\end{equation}
Thus
\begin{equation}\label{4.93}
\varphi=k_1Q^4\exp(\int\frac{c\md t}{g})
\end{equation} with $k_1\in\mathbb{R}$. Together with \eqref{4.83}, one
gets
\begin{equation}\label{4.94}
\mu=k_2Q^{-1}g^{-1}\exp(\int\frac{-2c\md t}{g})
\end{equation} with $k_2\in\mathbb{R}$. Hence
\begin{eqnarray}\label{4.95}
a_{11}&=&-\frac{Q'}{4Q}-\frac{c}{g}\nonumber\\
      &=&-\frac{\alpha'\alpha''+\beta'\sin\alpha\ (\beta'\sin\alpha)'}{2((\alpha')^2+(\beta'\sin\alpha)^2)}-\frac{c}{g}.
\end{eqnarray}

\begin{thm}
  Let $\alpha$ and $g$ be arbitrary functions of $t$. The functions $\beta$,
  $Q$, $\varphi$, $a_{22}$, $a_{33}$, $a_{23}$, $\mu$, $\xi_r$, $\psi_r$, $\phi_s$ and $\psi_s$
  are given by \eqref{4.89}, \eqref{4.91}, \eqref{4.79}, \eqref{4.80}, \eqref{4.81},
  \eqref{4.93}, \eqref{4.94}, \eqref{4.62}, \eqref{4.63} and
  \eqref{4.64}, respectively. Then we have the following solutions of the MHD
  equations \eqref{M1}-\eqref{M4}:

  \begin{equation}\label{4.96}
  \left\{\begin{aligned}
       u=&-((\frac{Q'}{4Q}-\frac{c}{g})\cos\alpha+\alpha'\sin\alpha)\x-(\alpha'\cos\alpha+a_{22}\sin\alpha)\y
                \\
         &-(\beta'\cos\alpha+a_{23})\z\sin\alpha-\varphi\mu Q\zeta_r\phi_s\cos\alpha\mp\varphi\mu Q^{\frac{1}{2}}\xi_r\psi_s\alpha'\sin\alpha,\\
       v=&(-(\frac{Q'}{4Q}-\frac{c}{g})\sin\alpha\cos\beta+\alpha'\cos\alpha\cos\beta-\beta'\sin\alpha\sin\beta)\x\\
         &+(-\alpha'\sin\alpha\cos\beta+a_{22}\cos\alpha\cos\beta+a_{23}\sin\beta)\y\\
         &+(-\beta'\sin^2\alpha\cos\beta+a_{23}\cos\alpha\cos\beta-a_{33}\sin\beta)\z\\
         &-\varphi\mu
         Q\zeta_r\phi_s\cos\beta\sin\alpha\pm(\alpha'\cos\alpha\cos\beta-\beta'\sin\alpha\sin\beta)\varphi\mu
         Q^{\frac{1}{2}}\xi_r\psi_s,\\
       w=&(-(\frac{Q'}{4Q}-\frac{c}{g})\sin\alpha\sin\beta+\alpha'\cos\alpha\sin\beta+\beta'\sin\alpha\cos\beta)\x\\
         &+(-\alpha'\sin\alpha\sin\beta+a_{22}\cos\alpha\sin\beta+a_{23}\cos\beta)\y\\
         &+(-\beta'\sin^2\alpha\sin\beta+a_{23}\cos\alpha\sin\beta-a_{33}\cos\beta)\z\\
         &-\varphi\mu
         Q\zeta_r\phi_s\sin\alpha\sin\beta\pm(\alpha'\cos\alpha\sin\beta+\beta'\sin\alpha\cos\beta)\varphi\mu
         Q^{\frac{1}{2}}\xi_r\psi_s,
  \end{aligned}\right.
  \end{equation}
\begin{equation}\label{4.97}
  \left\{\begin{aligned}
       H^1=&-\frac{2c}{Q}\sin\alpha((\alpha')^2-(\beta'\sin\alpha)^2)\y
                -\frac{4c}{Q}\alpha'\beta'\sin^2\alpha\z\\
         &-g\varphi\mu
         Q\zeta_r\phi_s\cos\alpha \mp g\varphi\mu\alpha'Q^{\frac{1}{2}}\xi_r\psi_s\sin\alpha, \\
       H^2=&(\frac{2c}{Q}\cos\alpha\cos\beta((\alpha')^2-(\beta'\sin\alpha)^2)-\frac{4c}{Q}\alpha'\beta'\sin\alpha\sin\beta)\y\\
         &+(\frac{4c}{Q}\alpha'\beta'\sin\alpha\cos\alpha\cos\beta+\frac{2c}{Q}\sin\beta((\alpha')^2-(\beta'\sin\alpha)^2))\z\\
         &-g\varphi\mu
         Q\zeta_r\phi_s\sin\alpha\cos\beta \pm(\alpha'\cos\alpha\cos\beta-\beta'\sin\alpha\sin\beta)g\varphi\mu
         Q^{\frac{1}{2}}\xi_r\psi_s,\\
       H^3=&(\frac{2c}{Q}\cos\alpha\sin\beta((\alpha')^2-(\beta'\sin\alpha)^2)+\frac{4c}{Q}\alpha'\beta'\sin\alpha\cos\beta)\y\\
         &+(\frac{4c}{Q}\alpha'\beta'\sin\alpha\cos\alpha\sin\beta-\frac{2c}{Q}\cos\beta((\alpha')^2-(\beta'\sin\alpha)^2))\z\\
         &-g\varphi\mu
         Q\zeta_r\phi_s\sin\alpha\sin\beta \pm(\alpha'\cos\alpha\sin\beta+\beta'\sin\alpha\cos\beta)g\varphi\mu
         Q^{\frac{1}{2}}\xi_r\psi_s,
  \end{aligned}\right.
  \end{equation}
  \begin{eqnarray}\label{4.98}
  p&=&-\rho((a_{11}'+a_{11}^2-Q)\frac{\x^2}{2}+(a_{22}'+a_{22}^2+a_{23}'-2a_{23}\beta'\cos\alpha-(\alpha')^2)\frac{\y^2}{2}\nonumber\\
   &&(a_{33}'+a_{33}^2+a_{23}'+2a_{23}\beta'\cos\alpha-(\beta'\sin\alpha)^2)\frac{\z^2}{2}\nonumber\\
   &&-(\alpha'+2\alpha'a_{22}+2a_{23}\beta'\sin\alpha -(\beta')^2\sin\alpha\cos\alpha)\x\y\nonumber\\
   &&-((\beta'\sin\alpha)'+2\alpha'a_{23}+\alpha'\beta'\cos\alpha)\x\z\nonumber\\
   &&+(a_{23}'-a_{11}a_{23}+(a_{22}-a_{33})\beta'\cos\alpha-\alpha'\beta'\sin\alpha)\y\z\nonumber\\
   &&+(\delta_{r,1}+4a\delta_{r,0})\varphi^2\mu^2Q^2\frac{\phi_r^2}{2}+(\delta_{s,1}+4b\delta_{s,0})\varphi^2\mu^2Q^2\frac{\xi_r^2}{2}-
   2\mu\xi_r\phi_s\nonumber\\
   &&\pm(-1)^r\mu\sqrt
   Q(\frac{g'}{g}+\frac{2c}{g}+\frac{Q'}{2Q})\zeta_r\psi_s
   \mp\varphi\mu\sqrt Q(\alpha'(\frac{\varphi'}{\varphi}+\frac{\alpha''}{\alpha'}
   +a_{22}\frac{\beta'\sin\alpha}{\alpha'}(a_{23}-\beta'\cos\alpha))\y\nonumber\\
   &&+\beta'\sin\alpha(\frac{\varphi'}{\varphi}+\frac{(\beta'\sin\alpha)'}{\beta'\sin\alpha}+a_{33}
   +\frac{\alpha'}{\beta'\sin\alpha}(a_{23}+\beta'\cos\alpha)\z)\xi_r\psi_s),
  \end{eqnarray} where $c$
  is an arbitrary constant.
\end{thm}

\vskip 1.0cm

Set
\begin{equation}\label{4.99}
\xi_r=\left\{\begin{array}{ll}
  be^{\gamma_1\y+\gamma_2\z}-ce^{-(\gamma_1\y+\gamma_2\z)} & \textrm{if
  } r=1,\\
  c\sin(\gamma_1\y+\gamma_2\z+b) &\textrm{if
  } r=0\\
\end{array}\right.
\end{equation} and
\begin{equation}\label{4.100}
\zeta_r=\left\{\begin{array}{ll}
  be^{\gamma_1\y+\gamma_2\z}+ce^{-(\gamma_1\y+\gamma_2\z)} & \textrm{if
  } r=1,\\
  c\cos(\gamma_1\y+\gamma_2\z+b) &\textrm{if
  } r=0;\\
\end{array}\right.
\end{equation}
where $\gamma_1$ and $\gamma_2$ are functions of $t$, and $b, c$ are
real constants.

Suppose that
\begin{equation}\label{4.101}
\begin{pmatrix}
  \U\\\V\\\W
\end{pmatrix}=A \begin{pmatrix}
  \x\\\y\\\z
\end{pmatrix}+ \begin{pmatrix}
  f_1-(g_1+(\gamma_1\sigma+\gamma_2\tau)\zeta_r)\\
  q_1+\sigma\xi_r\\
  s_1+\tau\xi_r
\end{pmatrix}
\end{equation}
and
\begin{equation}\label{4.102}
\begin{pmatrix}
  \h^1 \\ \h^2 \\ \h^3
\end{pmatrix}=B \begin{pmatrix}
  \x\\\y\\\z
\end{pmatrix}+ \begin{pmatrix}
  f_2-(g_2+(\gamma_1\sigma_1+\gamma_2\tau_1)\zeta_r)\\
  q_2+\sigma_1\xi_r\\
  s_2+\tau_1\xi_r
\end{pmatrix},
\end{equation}
where $A=(a_{i,j})_{3\times3}$ and $B=(b_{i,j})_{3\times3}$ are
$3\times 3$ matrices whose entries are functions of $t$; the
functions $f_j$, $q_j$, $s_j$, $\sigma$, $\tau$,$\sigma_1$ and
$\tau_1$ are also functions of $t$ to be determined.

The equations \eqref{4.16} become
\begin{eqnarray}\label{4.103}
&&b_{11}'\x+b_{12}'\y+b_{13}'\z+f_2'-(g_2'+(\gamma_1\sigma_1+\gamma_2\tau_1)'\x)\zeta_r\nonumber\\
&-&(-1)^r(\gamma_1'\y+\gamma_2'\z)(g_2+(\gamma_1\sigma_1+\gamma_2\tau_1)\x)\xi_r
+(\alpha'\y+\z\beta'\sin\alpha)(b_{11}-(\gamma_1\sigma_1+\gamma_2\tau_1)\zeta_r)\nonumber\\
&+&((a_{21}-\alpha')\x+a_{22}\y+(a_{23}+\beta'\cos\alpha)\z+q_1+\sigma\xi_r)
(b_{12}-(-1)^r\gamma_1(g_2+(\gamma_1\sigma_1+\gamma_2\tau_1)\x)\zeta_r)\nonumber\\
&-&((\beta'\sin\alpha-a_{31})\x+(\beta'\cos\alpha-a_{32})\y-a_{33}\z-s_1-\tau\xi_r)\nonumber\\
&\times&(b_{13}-(-1)^r\gamma_2(g_2+(\gamma_1\sigma_1+\gamma_2\tau_1)\x)\zeta_r)\nonumber\\
&+&((a_{22}+a_{33})+(\gamma_1\sigma+\gamma_2\tau)\zeta_r)(b_{11}\x+b_{12}\y+b_{13}\z+f_2-(g_2+(\gamma_1\sigma_1+\gamma_2\tau_1)\x)\zeta_r)\nonumber\\
&-&(\alpha'+a_{12}-(-1)^r\gamma_1(g_1+(\gamma_1\sigma+\gamma_2\tau)\x)\xi_r)
(b_{21}\x+b_{22}\y+b_{23}\z+q_2+\sigma_1\xi_r)\nonumber\\
&-&(\beta'\sin\alpha+a_{13}-(-1)^r\gamma_2(g_1+(\gamma_1\sigma+\gamma_2\tau)\x)\xi_r)(b_{31}\x+b_{32}\y+b_{33}\z+s_2+\sigma_2\xi_r)\nonumber\\
&-&(-b_{11}+(\gamma_1\sigma_1+\gamma_2\tau_1)\zeta_r)(a_{11}\x+a_{12}\y+a_{13}\z+f_1-(g_1+(\gamma_1\sigma+\gamma_2\tau)\x)\xi_r)\nonumber\\
&+&(-1)^r\eta(\gamma_1^2+\gamma_2^2)(g_2+(\gamma_1\sigma_1+\gamma_2\tau_1)\x)\zeta_r=0,
\end{eqnarray}
\begin{eqnarray}\label{4.104}
&&b_{21}'\x+b_{22}'\y+b_{23}'\z+q_2'+\sigma_1'\xi_r+(\gamma_1'\y+\gamma_2'\z)\sigma_1\xi_r
\nonumber\\
&+&b_{21}(a_{11}\x+(a_{12}+\alpha')\y+(a_{13}+\beta'\sin\alpha)\z+f_1-(g_1+(\gamma_1\sigma+\gamma_2\tau)\x)\xi_r)\nonumber\\
&+&(b_{23}+\gamma_2\sigma_1\xi_r)((a_{31}-\beta'\sin\alpha)\x+(a_{32}-\beta'\cos\alpha)\y+a_{33}\z+s_1+\tau\xi_r)\nonumber\\
&+&(\alpha'-a_{21})(b_{11}\x+b_{12}\y+b_{13}\z+f_2-(g_2+(\gamma_1\sigma_1+\gamma_2\tau_1)\x)\xi_r)\nonumber\\
&-&(a_{22}+\gamma_1\sigma\xi_r)(b_{21}\x+b_{22}\y+b_{23}\z+q_2+\sigma_1\xi_r)+(b_{22}+\gamma_1\sigma_1\xi_r)(-\alpha'\x+\z\beta'\cos\alpha)\nonumber\\
&-&(a_{23}+\beta'\cos\alpha+\gamma_2\sigma\xi_r)(b_{31}\x+b_{32}\y+b_{33}\z+s_2+\tau_1\xi_r)\nonumber\\
&+&(b_{22}+\gamma_1\sigma_1\xi_r)(a_{21}\x+a_{22}\y+a_{23}\z+q_1+\sigma\xi_r)-(-1)^r\eta(\gamma_1^2+\gamma_2^2)\sigma_1\xi_r=0
\end{eqnarray} and
\begin{eqnarray}\label{4.105}
&&b_{31}'\x+b_{32}'\y+b_{33}'\z+s_2'+\tau_1'\xi_r+(\gamma_1'\y+\gamma_2'\z)\tau_1\xi_r
\nonumber\\
&+&b_{31}(a_{11}\x+(a_{12}+\alpha')\y+(a_{13}+\beta'\sin\alpha)\z+f_1-(g_1+(\gamma_1\sigma+\gamma_2\tau)\x)\xi_r)\nonumber\\
&+&(b_{32}+\gamma_1\tau_1\xi_r)((a_{21}-\alpha')\x+a_{22}\y+(a_{23}+\beta'\cos\alpha)\z+q_1+\sigma\xi_r)\nonumber\\
&+&(b_{33}+\gamma_2\tau_1\xi_r)((a_{31}-\beta'\sin\alpha)\x+(a_{32}-\beta'\cos\alpha)\y+a_{33}\z+s_1+\tau\xi_r)\nonumber\\
&+&(\beta'\sin\alpha-a_{31})(b_{11}\x+b_{12}\y+b_{13}\z+f_2-(g_2+(\gamma_1\sigma_1+\gamma_2\tau_1)\x)\xi_r)\nonumber\\
&+&(\beta'\cos\alpha-a_{32}-\gamma_1\tau\xi_r)(b_{21}\x+b_{22}\y+b_{23}\z+q_2+\sigma_1\xi_r)\nonumber\\
&-&(a_{33}+\gamma_2\tau\xi_r)(b_{31}\x+b_{32}\y+b_{33}\z+s_2+\tau_1\xi_r)-(-1)^r\eta(\gamma_1^2+\gamma_2^2)\tau_1\xi_r=0.\nonumber\\
\end{eqnarray}
The coefficients of $\xi_r\zeta_r$ and $\xi_r^2$ say that
\begin{equation}\label{4.106}
\left\{\begin{aligned}
  \tau_1\sigma-\tau\sigma_1&=0,\\
  (\gamma_1\sigma_1+\gamma_2\tau_1)g_1&=(\gamma_1\sigma+\gamma_2\tau)g_2.
\end{aligned}\right.
\end{equation} Moreover, the coefficients of $\xi_r\zeta_r$ in $\Phi_{1\z}=\Phi_{3\x}$
yield
\begin{equation}\label{4.107}
\gamma_1\tau=\gamma_2\sigma.
\end{equation}
The above two equations suggest that
\begin{equation}\label{4.108}
\left\{ \begin{aligned}
\sigma_1&=h\sigma,\\
\tau_1&=h\tau,
\end{aligned}\right.
\end{equation} where
\begin{equation}\label{4.109}
h=\frac{g_2}{g_1}. \end{equation} Motivated by Theorem 4.4, we
assume that
\begin{equation}\label{4.110}
a_{21}=-a_{12}=\alpha',\ \ \ \ a_{31}=-a_{13}=\beta'\sin\alpha.
\end{equation}
 The coefficients of $\xi_r$ in $\Phi_{1\y}=\Phi_{2\x}$
and the coefficients of $\y$ in \eqref{4.103} show that
\begin{equation}\label{4.111}
b_{11}=b_{12}=b_{13}=0.
\end{equation}
The coefficients of $\xi_r$ and $\zeta_r$ in
\eqref{4.103}-\eqref{4.105} yield
\begin{equation}\label{4.112}
b_{21}=b_{31}=0.
\end{equation}
Now
\begin{eqnarray}\label{4.113}
R_1&=&(a_{11}'+a_{11}^2+(\beta'\sin\alpha)^2)\x-(\alpha''+2\alpha'a_{22}-2a_{32}\beta'\sin\alpha
+(\beta')^2\sin\alpha\cos\alpha)\y\nonumber\\
&&+(-(\beta'\sin\alpha)'-\alpha'(a_{23}+\beta'\cos\alpha)-2a_{33}\beta'\sin\alpha-\alpha'a_{23})\z\nonumber\\
&&+\{-2\alpha'\sigma-2\beta'\tau\sin\alpha-(-1)^r(g_1+(\gamma_1\sigma+\gamma_2\tau)\x)
[(\gamma_1'+\gamma_1a_{22}+\gamma_2(a_{32}-\beta'\cos\alpha))\y\nonumber\\
&&+(\gamma_2'+\gamma_1(a_{23}+\beta'\cos\alpha)-2a_{33}\beta'\sin\alpha
)\z+\gamma_1q_1+\gamma_2s_1]\}\xi_r+\{-(g_1'+(\gamma_1\sigma+\gamma_2\tau)'\x)\nonumber\\
&&-a_{11}(g_1+(\gamma_1\sigma+\gamma_2\tau)\x)-(a_{11}\x+f_1)(\gamma_1\sigma+\gamma_2\tau)\nonumber\\
&&+(-1)^r\nu(\gamma_1^2+\gamma_2^2)(g_1+(\gamma_1\sigma+\gamma_2\tau)\x)\}\zeta_r\nonumber\\
&&+(4bc\delta_{r,0}+c^2\delta_{r,1})(\gamma_1\sigma+\gamma_2\tau)(g_1+(\gamma_1\sigma+\gamma_2\tau)\x)\nonumber\\
&&+f_1'+a_{11}f_1-2\alpha'q_1-2s_1\beta'\sin\alpha,
\end{eqnarray}
\begin{eqnarray}\label{4.114}
R_2&=&(\alpha''+2\alpha'a_{11}-(\beta')^2\sin\alpha\cos\alpha)\x\nonumber\\
   &&+(a_{22}'+a_{22}^2+a_{23}(a_{32}-\beta'\cos\alpha)-(\alpha')^2-a_{32}\beta'\cos\alpha
      )\y\nonumber\\
   &&+(a_{23}'+a_{22}(a_{23}+\beta'\cos\alpha)+a_{23}a_{33}-\alpha'\beta'\sin\alpha-a_{33}\beta'\cos\alpha
         )\z\nonumber\\
   &&+(\sigma'+a_{22}\sigma+(a_{23}-\beta'\cos\alpha)\tau-(-1)^r\nu(\gamma_1^2+\gamma_2^2)\sigma)\xi_r\nonumber\\
   &&+\{-2\alpha'(g_1+(\gamma_1\sigma+\gamma_2\tau)\x)+\sigma[(\gamma_1'+\gamma_1a_{22}+\gamma_2(a_{32}-\beta'\cos\alpha))\y\nonumber\\
   &&+(\gamma_2'+\gamma_1(a_{23}+\beta'\cos\alpha)-2a_{33}\beta'\sin\alpha
)\z+\gamma_1q_1+\gamma_2s_1]\}\zeta_r\nonumber\\
&&+\sigma(\gamma_1\sigma+\gamma_2\tau)\xi_r\zeta_r+q_1'+a_{22}q_1+(a_{23}-\beta'\cos\alpha)s_1,
\end{eqnarray}
\begin{eqnarray}\label{4.115}
R_3&=&((\beta'\sin\alpha)'+2a_{11}\beta'\sin\alpha +\alpha'\beta'\cos\alpha)\x\nonumber\\
   &&+(a_{32}'+a_{32}a_{22}+a_{33}(a_{32}-\beta'\cos\alpha)-\alpha'\beta'\sin\alpha-a_{22}\beta'\cos\alpha
      )\y\nonumber\\
   &&+(a_{33}'+a_{32}(a_{23}+\beta'\cos\alpha)+a_{33}^2-(\beta'\sin\alpha)^2+a_{23}\beta'\cos\alpha
         )\z\nonumber\\
   &&+(\tau'+a_{33}\tau+(a_{32}+\beta'\cos\alpha)\sigma-(-1)^r\nu(\gamma_1^2+\gamma_2^2)\tau)\xi_r\nonumber\\
   &&+\{-2(g_1+(\gamma_1\sigma+\gamma_2\tau)\x)\beta'\sin\alpha+\tau[(\gamma_1'+\gamma_1a_{22}+\gamma_2(a_{32}-\beta'\cos\alpha))\y\nonumber\\
   &&+(\gamma_2'+\gamma_1(a_{23}+\beta'\cos\alpha)-2a_{33}\beta'\sin\alpha
)\z+\gamma_1q_1+\gamma_2s_1]\}\zeta_r\nonumber\\
&&+\tau(\gamma_1\sigma+\gamma_2\tau)\xi_r\zeta_r+s_1'+a_{33}s_1+(a_{32}+\beta'\cos\alpha)q_1+2f_1\beta'\sin\alpha
,
\end{eqnarray}
\begin{eqnarray}\label{4.116}
   G_1&=&(-1)^rh^2(\gamma_1\sigma+\gamma_2\tau)(g_1+(\gamma_1\sigma+\gamma_2\tau)\x)\xi_r^2\nonumber\\
   &&+(-1)^rh(g_1+(\gamma_1\sigma+\gamma_2\tau)\x)(\gamma_1b_{23}\z+\gamma_2b_{32}\y+\gamma_1q_2+\gamma_2s_2)\xi_r,
\end{eqnarray}
\begin{eqnarray}\label{4.117}
G_2&=&b_{32}(b_{32}-b_{23})\y+h^2(-1)^r\gamma_1(g_1+(\gamma_1\sigma+\gamma_2\tau)\x)^2\xi_r\zeta_r\nonumber\\
&&+h((b_{32}-b_{23})\tau-(-1)^rf_2\gamma_1(g_1+(\gamma_1\sigma+\gamma_2\tau)\x))\xi_r
+(b_{32}-b_{23})s_2,
\end{eqnarray}
\begin{eqnarray}\label{4.118}
G_3&=&b_{23}(b_{32}-b_{23})\z+h^2(-1)^r\gamma_2(g_1+(\gamma_1\sigma+\gamma_2\tau)\x)^2\xi_r\zeta_r\nonumber\\
&&-h((b_{32}-b_{23})\sigma-(-1)^rf_2\gamma_1(g_1+(\gamma_1\sigma+\gamma_2\tau)\x))\xi_r.
\end{eqnarray}
By \eqref{4.18}, we have that
\begin{equation}\label{4.119}
\left\{\begin{aligned}
  a_{33}&=\frac{\alpha''}{\alpha}-\frac{(\beta')^2\sin\alpha\cos\alpha}{\alpha'}+\frac{\beta'\sin\alpha\;
  a_{32}}{\alpha'},\\
  a_{22}&=\frac{(\beta'\sin\alpha)'}{\beta'\sin\alpha}+\frac{\alpha'\cos\alpha}{\sin\alpha}+\frac{\alpha'a_{23}}
  {\beta'\sin\alpha}
\end{aligned}\right.
\end{equation} and
\begin{equation}\label{4.120}
(a_{23}-a_{32})a_{11}=a_{23}'-a_{32}'.
\end{equation}
The coefficients of $\zeta_r$ in $\Phi_{1y}=\Phi_{2x}$ say that
\begin{equation}\label{4.121}
\left\{\begin{aligned}
 b_{32}&=-\frac{\gamma_1'+\gamma_1a_{22}+\gamma_2(a_{32}-\beta'\cos\alpha)}{\mu_0h\gamma_2},\\
 b_{23}&=-\frac{\gamma_2'+\gamma_1(a_{23}+\beta'\cos\alpha)+\gamma_2a_{33}}{\mu_0h\gamma_1}.
\end{aligned}\right.
\end{equation}
But the coefficients of $\xi$ in \eqref{4.103} yield
\begin{equation}\label{4.122}
\left\{\begin{aligned}
    b_{32}&=\frac{h(\gamma_1'+\gamma_1a_{22}+\gamma_2(a_{32}-\beta'\cos\alpha))}{\gamma_2},\\
    b_{23}&=\frac{h(\gamma_2'+\gamma_1(a_{23}+\beta'\cos\alpha)+\gamma_2a_{33})}{\gamma_1}.
\end{aligned} \right.
\end{equation}
Thus
\begin{equation}\label{4.123}
b_{23}=b_{32}=0
\end{equation} and
\begin{equation}\label{4.124}
\left\{\begin{aligned}
\gamma_1'+\gamma_1a_{22}+\gamma_2(a_{32}-\beta'\cos\alpha)&=0,\\
\gamma_2'+\gamma_1(a_{23}+\beta'\cos\alpha)+\gamma_2a_{33}&=0.
\end{aligned}\right.
\end{equation}
Again by \eqref{4.18}, we have that
\begin{equation}\label{4.125}\left\{\begin{aligned}
&\frac{(\gamma_1\sigma+\gamma_2\tau)'}{\gamma_1\sigma+\gamma_2\tau}+2a_{11}-(-1)^r\nu(\gamma_1^2+\gamma_2^2)=0,\\
&\frac{g_1'}{g_1}+a_{11}+\frac{f_1(\gamma_1\sigma+\gamma_2\tau)}{g_1}-(-1)^r\nu(\gamma_1^2+\gamma_2^2)
+\frac{(-1)^r\mu_0f_2h(\gamma_1\sigma+\gamma_2\tau)}{g_1}=0,\end{aligned}\right.
\end{equation}
\begin{equation}\label{4.126}
\gamma_1\beta'\sin\alpha=\alpha'\gamma_2
\end{equation}
and
\begin{equation}\label{4.127}
\gamma_1q_1+\gamma_2s_1+\mu_0h(\gamma_1q_2+\gamma_2s_2)=0.
\end{equation}
Since
\begin{equation}\label{4.128}
h(\gamma_1q_1+\gamma_2s_1)+\gamma_1q_2+\gamma_2s_2=0,
\end{equation} comparing with \eqref{4.103}, we get
\begin{equation}\label{4.129}
\gamma_1q_1+\gamma_2s_1=\gamma_1q_2+\gamma_2s_2=0.
\end{equation}
Furthermore, by \eqref{4.107} and \eqref{4.126}, we can write
\begin{equation}\label{4.130}
\left\{\begin{aligned}
 \gamma_1&=\varphi\alpha'\\
 \gamma_2&=\varphi\beta'\sin\alpha
\end{aligned}\right.\ \ \ \ \textrm{and}\ \ \ \ \left\{\begin{aligned}
 \sigma&=\mu\alpha'\\
 \tau&=\mu\beta'\sin\alpha
\end{aligned}\right.
\end{equation} for some functions $\varphi$ and $\mu$.

From \eqref{4.103}-\eqref{4.105}, we have that
\begin{equation}\label{4.131}
\left\{\begin{aligned}
 &(h(\gamma_1\sigma+\gamma_2\tau))'-(-1)^r\eta
h(\gamma_1^2+\gamma_2^2)g_1=0\\
&(hg_1)'-ha_{11}g_1+hf_1(\gamma_1\sigma+\gamma_2\tau)-f_2(\gamma_1\sigma+\gamma_2\tau)-\eta(-1)^r\eta
h(\gamma_1^2+\gamma_2^2)g_1=0,
\end{aligned}\right.
\end{equation}
\begin{equation}\label{4.132}
\left\{\begin{aligned}
  (\sigma h)'-a_{12}h\sigma-h\tau(a_{23}+\beta'\cos\alpha)-(-1)^r\eta
h(\gamma_1^2+\gamma_2^2)\sigma&=0\\
  (\tau h)'+(\beta'\cos\alpha-a_{32})h\sigma-a_{33}h\tau-(-1)^r\eta
h(\gamma_1^2+\gamma_2^2)\tau&=0
\end{aligned}\right.
\end{equation} and
\begin{equation}\label{4.133}
\left\{\begin{aligned}
   &f_2'-a_{11}f_2=0\\
   &q_2'-a_{22}q_2-(a_{23}+\beta'\cos\alpha)s_2=0\\
   &s_2'+(\beta'\cos\alpha-a_{32})q_2-a_{33}s_2=0.
\end{aligned}\right.
\end{equation}
For simplicity, we assume that
\begin{equation}\label{4.134}
f_2=q_2=s_2=0.
\end{equation}
Then by \eqref{4.119} and \eqref{4.124}, one has
\begin{equation}\label{4.135}
\left\{\begin{aligned}
  a_{22}&=\frac{(\beta'\sin\alpha)'}{\beta'\sin\alpha}+\frac{\alpha'\beta'\cos\alpha}{\beta'\sin\alpha}
  +\frac{\alpha'a_{23}}{\beta'\sin\alpha}\\
  a_{33}&=-\frac{\varphi'}{\varphi}-\frac{(\beta'\sin\alpha)'}{\beta'\sin\alpha}-\frac{\alpha'\beta'\cos\alpha}{\beta'\sin\alpha}
  -\frac{\alpha'a_{23}}{\beta'\sin\alpha}\\
  a_{32}&=-\frac{\alpha'}{\beta'\sin\alpha}(\frac{\varphi'}{\varphi}+\frac{\alpha''}{\alpha}
  +\frac{(\beta'\sin\alpha)'}{\beta'\sin\alpha}+\frac{\alpha'\beta'\cos\alpha}{\beta'\sin\alpha})+\beta'\cos\alpha
  -\frac{(\alpha')^2a_{23}}{(\beta'\sin\alpha)^2}.
\end{aligned}\right.
\end{equation}
Thus
\begin{equation}\label{4.136}
a_{11}=\frac{\varphi'}{\varphi}.
\end{equation}
The system \eqref{4.125} says that
\begin{equation}\label{4.137}
3\frac{\varphi'}{\varphi}+\frac{\mu'}{\mu}+\frac{Q'}{Q}=(-1)^r\nu\varphi
Q,
\end{equation} where
\begin{equation}\label{4.138}
Q=(\alpha')^2+(\beta'\sin\alpha)^2.
\end{equation} Then we get
\begin{equation}\label{4.139}
\mu=\frac{c_1}{\varphi^3Q}\exp(\int(-1)^r\nu\varphi Q\md t)
\end{equation}
for some constants $c_1$.

We consider the special case of \eqref{4.135}, in which
\begin{equation}\label{4.140}
a_{23}=a_{32}.
\end{equation}
Hence we have
\begin{equation}\label{4.141}
a_{23}=-\frac{\alpha'\beta'\sin\alpha}{Q}(\frac{\varphi'}{\varphi}+\frac{\alpha''}{\alpha}
  +\frac{(\beta'\sin\alpha)'}{\beta'\sin\alpha}+\frac{\alpha'\beta'\cos\alpha}{\beta'\sin\alpha}-\frac{(\beta')^2\sin\alpha\cos\alpha}{\alpha'}).
\end{equation}
One shows that the system \eqref{4.131} is compatible by
\eqref{4.125}. Moreover, the equation $\eqref{4.132}_1$ is
compatible with \eqref{4.131} if and only if
\begin{equation}\label{4.142}
\alpha''\beta'\sin\alpha-\alpha'\beta''\sin\alpha-2(\alpha')^2\beta'\cos\alpha-(\beta')^3\sin^2\alpha\
\cos\alpha=0,
\end{equation} i.e.
\begin{equation}\label{4.143}
\beta=\pm\int\frac{\md \alpha}{\sin^2\alpha\
\sqrt{d-\sin^{-2}\alpha}}+d_0
\end{equation} for some constants $d_0$ and $d>1$. In this case, one shows the
system \eqref{4.132} is compatible. Thus one gets that
\begin{equation}\label{4.144}
h=c_2\varphi^2\exp((-1)^r\int (\eta-\nu)\varphi Q\md t),
\end{equation} where $c_2$ is a real constant.

Write
\begin{equation}\label{4.145}
\left\{ \begin{aligned}
  s_1&=\lambda\alpha',\\
  q_1&=\lambda\beta'\sin\alpha.
\end{aligned}\right.
\end{equation} Then we have
\begin{eqnarray}\label{4.148}
p&=&-\rho(-(-1)^r(c_2\varphi^2\exp((-1)^r\int(\eta-\nu)\varphi Q\md
t))^2\mu_0((g_1+\varphi\mu Q\x)^2\xi_r^2)\nonumber\\
&+&(\frac{\varphi''}{\varphi}-Q)\frac{\x^2}{2}+(a_{22}'+a_{22}^2+a_{23}^2-2a_{23}\beta'\cos\alpha
-(\alpha')^2)\frac{\y^2}{2}\nonumber\\
&+&(a_{33}'+a_{23}'+a_{33}^2+2a_{23}\beta'\cos\alpha-(\beta'\sin\alpha)^2)\frac{\z^2}{2}
+(\alpha''+2\alpha'\frac{\varphi'}{\varphi}-(\beta')^2\sin\alpha\cos\alpha)\x\y\nonumber\\
&+&((\beta'\sin\alpha)'+2\frac{\varphi'}{\varphi}\beta'\sin\alpha+\alpha'\beta'\cos\alpha)\x\z\nonumber\\
&+&(a_{23}-a_{23}\frac{\varphi'}{\varphi}+(a_{22}-a_{33})\beta'\cos\alpha-\alpha'\beta'\cos\alpha)\y\z\nonumber\\
&+&\frac{1}{2}(4bc\delta_{r,0}+c^2\delta_{r,1})(g_1+\varphi\mu
Q\x)^2+(f_1'+\frac{\varphi'f_1}{f_1}-4\lambda\alpha'\beta'\sin\alpha)\x\nonumber\\
&+&((\lambda\beta\sin\alpha)'+a_{22}\lambda\beta\sin\alpha+(a_{23}-\beta'\cos\alpha)(\lambda\alpha'))\y\nonumber\\
&+&((\lambda\alpha')'+2f_1\beta'\sin\alpha
+(a_{23}+\beta'\cos\alpha)\lambda\beta'\sin\alpha+a_{33}\lambda\alpha')\z\nonumber\\
&-&2\mu
Q\x\xi_r-2\frac{\xi_r}{\varphi}+(-1)^r\frac{\mu}{\varphi}(\frac{\mu'}{\mu}-\frac{\varphi'}{\varphi}-(-1)^r\nu\varphi^2Q)\zeta_r)
\end{eqnarray} modulo the transformation in \eqref{T5}, where $c_2$ is an arbitrary constant.

\begin{thm}
Let $\alpha$, $\varphi$, $g_1$, $\lambda$, $f_1$ be functions of
$t$. The functions $\mu$, $\xi_r$, $\zeta_r$, $\beta$, $Q$,
$a_{22}$, $a_{33}$, $a_{23}$ are given by \eqref{4.139},
\eqref{4.99}, \eqref{4.100}, \eqref{4.143}, \eqref{4.138},
\eqref{4.135} and \eqref{4.141}, respectively. Then we have the
following solutions of the MHD equations \eqref{M1}-\eqref{M4}:
\begin{equation}\label{4.146}
\left\{\begin{aligned}
 u=&(\frac{\varphi'}{\varphi}\cos\alpha-\alpha'\sin\alpha)\x-(\alpha'\cos\alpha+a_{22}\sin\alpha)\y
 -(\beta'\cos\alpha-a_{23})\z\sin\alpha\\
 &+f_1\cos\alpha-\lambda\beta'\sin^2\alpha-(g_1+\varphi\mu
 Q\x)\zeta_r\cos\alpha-\mu\xi_r\alpha'\sin\alpha,\\
 v=&(\frac{\varphi'}{\varphi}\sin\alpha\cos\beta+\alpha'\cos\alpha\cos\beta-\beta'\sin\alpha\sin\beta)\x\\
 &+(-\alpha'\sin\alpha\cos\beta+a_{22}\cos\alpha\cos\beta-a_{23}\sin\beta)\y\\
 &+(-\beta'\sin^2\alpha\cos\beta+a_{23}\cos\alpha\cos\beta-a_{33}\sin\beta)\z\\
 &+f_1\sin\alpha\cos\beta+\lambda\beta'\sin\alpha\cos\alpha\cos\beta-\lambda\alpha'\sin\beta\\
 &-(g_1+\varphi\mu
 Q\x)\zeta_r\sin\alpha\cos\beta+\mu\alpha'\xi_r\cos\alpha\cos\beta-\mu\beta'\xi_r\sin\alpha\sin\beta,\\
 w=&(\frac{\varphi'}{\varphi}\sin\alpha\sin\beta+\alpha'\cos\alpha\sin\beta+\beta'\sin\alpha\cos\beta)\x\\
 &+(-\alpha'\sin\alpha\sin\beta+a_{22}\cos\alpha\sin\beta+a_{23}\cos\beta)\y\\
 &+(-\beta'\sin^2\alpha\sin\beta+a_{23}\cos\alpha\sin\beta+a_{33}\cos\beta)\z\\
 &+f_1\sin\alpha\sin\beta+\lambda\beta'\sin\alpha\cos\alpha\sin\beta+\lambda\alpha'\cos\beta\\
 &-(g_1+\varphi\mu
 Q\x)\zeta_r\sin\alpha\sin\beta+\mu\alpha'\xi_r\cos\alpha\sin\beta+\mu\beta'\xi_r\sin\alpha\cos\beta,
\end{aligned}\right.
\end{equation}
\begin{equation}\label{4.147}
\left\{\begin{aligned}
   H^1=&-c_2\varphi^2\exp((-1)^r\int(\eta-\nu)\varphi Q\md t)((g_1+\varphi\mu
 Q\x)\zeta_r\cos\alpha+\mu\alpha'\xi_r\sin\alpha),\\
   H^2=&c_2\varphi^2\exp((-1)^r\int(\eta-\nu)\varphi Q\md t)(-(g_1+\varphi\mu
 Q\x)\zeta_r\sin\alpha\cos\beta\\
 &+\mu\alpha'\xi_r\cos\alpha\cos\beta-\mu\beta'\xi_r\sin\alpha\sin\beta),\\
   H^3=&c_2\varphi^2\exp((-1)^r\int(\eta-\nu)\varphi Q\md t)(-(g_1+\varphi\mu
 Q\x)\zeta_r\sin\alpha\sin\beta\\&+\mu\alpha'\xi_r\cos\alpha\sin\beta+\mu\beta'\xi_r\sin\alpha\cos\beta)
\end{aligned}\right.
\end{equation} and $p$ is given by \eqref{4.148}.
\end{thm}

\vskip 1.0cm

Set
\begin{equation}\label{4.149}
\phi_r=\left\{\begin{array}{ll}
   e^{a_1\y}-ae^{-a_1\y} & \textrm{if } r=0,\\
   \sin(a_1\y) & \textrm{if } r=1,
\end{array}\right.\ \ \ \ \psi_r=\left\{\begin{array}{ll}
   e^{a_1\y}+ae^{-a_1\y} & \textrm{if } r=0,\\
   \cos(a_1\y) & \textrm{if } r=1,
\end{array}\right.
\end{equation}
\begin{equation}\label{4.150}
\xi_r=\left\{\begin{array}{ll}
   be^{a_1\y}-ce^{-a_1\y} & \textrm{if } r=0,\\
   c\sin(a_1\y+b) & \textrm{if } r=1,
\end{array}\right.\ \ \ \ \zeta_r=\left\{\begin{array}{ll}
   be^{a_1\y}+ce^{-a_1\y} & \textrm{if } r=0,\\
   c\cos(a_1\y+b) & \textrm{if } r=1
\end{array}\right.
\end{equation}and
\begin{equation}\label{4.151}
\theta_r=\left\{\begin{array}{ll}
   b_1e^{a_1\y}-c_1e^{-a_1\y} & \textrm{if } r=0,\\
   c_1\sin(a_1\y+b_1) & \textrm{if } r=1,
\end{array}\right.\ \ \ \ \varepsilon_r=\left\{\begin{array}{ll}
   b_1e^{a_1\y}+c_1e^{-a_1\y} & \textrm{if } r=0,\\
   c_1\cos(a_1\y+b_1) & \textrm{if } r=1,
\end{array}\right.
\end{equation} where $a_1$ is
a functions of $t$ and $a$, $b$, $b_1$, $c$, $c_1$ are real
constants .

We assume that
\begin{equation}\label{4.152}
\left\{\begin{aligned}
   \U&=a_{11}\x+a_{12}\y+a_{13}\z+a_1f(\theta_r+a_2\y\phi_r-\x\zeta_r-a_1a_2\x^2\psi_r),\\
   \V&=a_{21}\x+a_{22}\y+a_{23}\z+f(\xi_r+2a_1a_2\x\phi_r),\\
   \W&=a_{31}\x+a_{32}\y+a_{33}\z
\end{aligned}\right.
\end{equation} and
\begin{equation}\label{4.153}
\left\{\begin{aligned}
   \h^1&=b_{11}\x+b_{12}\y+b_{13}\z+a_1g(\theta_r+b_2\y\phi_r-\x\zeta_r-a_1b_2\x^2\psi_r),\\
   \V&=b_{21}\x+b_{22}\y+b_{23}\z+g(\xi_r+2a_1b_2\x\phi_r),\\
   \W&=b_{31}\x+b_{32}\y+b_{33}\z,
\end{aligned}\right.
\end{equation} where $a_{ij}$, $b_{ij}$, $a_2(\neq0)$, $b_2(\neq0)$, $f(\neq0)$ and $g(\neq0)$
are functions of $t$, and
\begin{equation}\label{4.154}
\left\{\begin{aligned}
    a_{11}+a_{22}+a_{33}&=0,\\
    b_{11}+b_{22}+b_{33}&=0.\\
\end{aligned}\right.
\end{equation}

Substituting \eqref{4.152}-\eqref{4.154} into \eqref{4.16} and
\eqref{4.18}, we get that
\begin{eqnarray}\label{4.155}
&&R_1=a_{11}'\x+a_{12}'\y+a_{13}'\z+(a_1f)'(\theta_r+a_2\y\phi_r-\x\zeta_r-a_1a_2\x^2\psi_r)\nonumber\\
&+&a_1f(a_1'\y\varepsilon_r+a_2'\y\phi_r+a_1'a_2\y^2\psi_r-(-1)^ra_1'\x\y\xi_r-(a_1a_2)'\x\psi_r-(-1)^ra_1'a_1a_2\x^2\y\phi_r)\nonumber\\
&+&(a_{11}-a_1f(\zeta_r+2a_1a_2\x\psi_r))(a_{11}\x+(a_12+\alpha')\y+(a_{13}+\beta'\sin\alpha)\z\nonumber\\
&+&a_1f(\theta_r+a_2\y\phi_r-\x\zeta_r-a_1a_2\x^2\psi_r))\nonumber\\
&+&(a_{12}+a_1f(a_1\varepsilon_r+a_2\phi_r+a_1a_2\y\psi_r-(-1)^ra_1\x\xi_r-(-1)^ra_1^2a_2\x^2\phi_r))\nonumber\\
&\times&((a_{21}-\alpha')\x+a_{22}\y+(a_{23}+\beta'\cos\alpha)\z+f(\xi_r+2a_1a_2\x\phi_r))\nonumber\\
&+&a_{13}((a_{31}-\beta'\sin\alpha)\x+(a_{32}-\beta'\cos\alpha)\y+a_{33}\z)\nonumber\\
&-&\alpha'(a_{21}\x+a_{22}\y+a_{23}\z+f(\xi_r+2a_1a_2\x\phi_r))-(a_{31}\x+a_{32}\y+a_{33}\z)\beta'\sin\alpha\nonumber\\
&-&(-1)^r\nu
a_1^3f(\theta_r+a_2\y\phi_r-\x\zeta_r-a_1a_2\x^2\psi_r),
\end{eqnarray}
\begin{eqnarray}\label{4.156}
 &&G_1=(b_{21}\x+b_{22}\y+b_{23}\z+g(\xi_r+2a_1b_2\x\phi_r))\nonumber\\
 &\times&(b_{21}-b_{12}+a_1g(-a_1\varepsilon_r+b_2\phi_r-a_1b_2\y\psi_r+(-1)^ra_1(\x\xi_r+a_1b_2\x^2\phi_r))\nonumber\\
 &-&(b_{13}-b_{31})(b_{31}\x+b_{32}\y+b_{33}\z),
\end{eqnarray}
\begin{eqnarray}\label{4.157}
&&R_2=a_{21}'\x+a_{22}'\y+a_{23}'\z+f'(\xi_r+2a_1a_2\x\phi_r)\nonumber\\&+&f(a_1'\y\zeta_r+2(a_1a_2)'\x\phi_r+2a_1'a_1a_2\x\y\psi_r)\nonumber\\
&+&(a_{21}+2a_1a_2f\phi_r)(a_{11}\x+(a_{12}+\alpha')\y+(a_{13}+\beta'\sin\alpha)\z\nonumber\\
&+&a_1f(\theta_r+a_2\y\phi_r-\x\zeta_r-a_1a_2\x^2\psi_r))\nonumber\\
&+&(a_{22}+a_1f(\zeta_r+2a_1a_2\x\psi_r))((a_{21}-\alpha')\x+a_{22}\y+(a_{23}+\beta'\cos\alpha)\z+f(\xi_r+2a_1a_2\x\phi_r))\nonumber\\
&+&a_{23}((a_{31}-\beta'\sin\alpha)\x+(a_{32}-\beta'\cos\alpha)\y+a_{33}\z)\nonumber\\
&+&\alpha'(a_{11}\x+a_{12}\y+a_{13}\z+a_1f(\theta_r+a_2\y\phi_r-\x\zeta_r-a_1a_2\x^2\psi_r))\nonumber\\
&-&(a_{31}\x+a_{32}\y+a_{33}\z)\beta'\cos\alpha-(-1)^r\nu
a_1^2f(\xi_r+2a_1a_2\x\phi_r),
\end{eqnarray}
\begin{eqnarray}\label{4.158}
&&G_2=(b_{32}-b_{23})(b_{31}\x+b_{32}\y+b_{33}\z)\nonumber\\
&-&(b_{21}-b_{12}+a_1g(-a_1\varepsilon_r+b_2\phi_r-a_1b_2\y\psi_r+(-1)^ra_1\x\xi_r+(-1)^ra_1^2b_2\x^2\phi_r))\nonumber\\
&\times&(b_{11}\x+b_{12}\y+b_{13}\z+a_1g(\theta_r+b_2\y\phi_r-\x\zeta_r-a_1b_2\x^2\psi_r)),
\end{eqnarray}
\begin{eqnarray}\label{4.159}
&&R_3=a_{31}'\x+a_{32}\y+a_{33}'\z+a_{31}(a_{11}\x+(a_{12}+\alpha')\y+(a_{13}+\beta'\sin\alpha)\z\nonumber\\
&+&a_1f(\theta_r+a_2\y\phi_r-\x\zeta_r-a_1a_2\x^2\psi_r))\nonumber\\
&+&a_{32}((a_{21}-\alpha')\x+a_{22}\y+(a_{23}+\beta'\cos\alpha)\z+f(\xi_r+2a_1a_2\x\phi_r))\nonumber\\
&+&a_{33}((a_{31}-\beta'\sin\alpha)\x+(a_{32}-\beta'\cos\alpha)\y+a_{33}\z)\nonumber\\
&+&(a_{11}\x+a_{12}\y+a_{13}\z+a_1f(\theta_r+a_2\y\phi_r-\x\zeta_r-a_1a_2\x^2\psi_r))\beta'\sin\alpha\nonumber\\
&+&(a_{11}\x+a_{12}\y+a_{13}\z+f(\xi_r+2a_1a_2\x\phi_r))\beta'\cos\alpha,
\end{eqnarray}
\begin{eqnarray}\label{4.160}
&&G_3=(b_{13}-b_{31})(b_{11}\x+b_{12}\y+b_{13}\z+a_1g(\theta_r+b_2\y\phi_r-\x\zeta_r-a_1b_2\x^2\psi_r))\nonumber\\
&-&(b_{32}-b_{23})(b_{21}\x+b_{22}\y+b_{23}\z+g(\xi_r+2a_1b_2\x\phi_r)),
\end{eqnarray}
\begin{eqnarray}\label{4.161}
&&b_{11}'\x+b_{12}'\y+b_{13}'\z+(a_1g)'(\theta_r+b_2\y\phi_r-\x\zeta_r-a_1b_2\x^2\psi_r)\nonumber\\
&+&a_1g(a_1'\y\varepsilon_r+b_2'\y\phi_r+a_1'b_2\y^2\psi_r-(-1)^ra_1'\x\y\xi_r-(a_1b_2)'\x\psi_r-(-1)^ra_1'a_1b_2\x^2\y\phi_r)\nonumber\\
&+&(\alpha'\y+\z\beta'\sin\alpha)(b_{11}-a_1g(\zeta_r+2a_1b_2\x\psi_r))\nonumber\\
&+&((a_{21}-\alpha')\x+a_{22}\y+(a_{23}+\beta'\cos\alpha)\z+f(\xi_r+2a_1a_2\x\phi_r))\nonumber\\
&\times&(b_{12}+a_1g(a_1\varepsilon_r+b_2\phi_r+a_1b_2\y\psi_r-(-1)^ra_1(\x\xi_r+a_1b_2\x^2\phi_r)))\nonumber\\
&+&b_{13}((a_{31}-\beta'\sin\alpha)\x+(a_{32}-\beta'\cos\alpha)\y+a_{23}\z)\nonumber\\
&+&(a_{22}+a_{33}+f(a_1\zeta_r+2a_1^2a_2\x\psi_r))\nonumber\\
&\times&(b_{11}\x+b_{12}\y+b_{13}\z+a_1g(\theta_r+b_2\y\phi_r-\x\zeta_r-a_1b_2\x^2\psi_r))\nonumber\\
&-&(\alpha'+a_{12}+a_1f(a_1\varepsilon_r+a_2\phi_r+a_1a_2\y\psi_r-(-1)^ra_1(\x\xi_r+a_1a_2\x^2\phi_r)))\nonumber\\
&\times&(b_{21}\x+b_{22}\y+b_{23}\z+g(\xi_r+2a_1b_2\x\phi_r))-(a_{13}+\beta'\sin\alpha)(b_{31}\x+b_{32}\y+b_{33}\z)\nonumber\\
&-&(a_{11}\x+a_{12}\y+a_{13}\z+a_1f(\theta_r+a_2\y\phi_r-\x\zeta_r-a_1a_2\x^2\psi_r))\nonumber\\
&\times&(b_{22}+b_{33}+g(a_1\zeta_r+2a_1^2b_2\x\psi_r))\nonumber\\
&-&(-1)^r\eta
a_1^3g(\theta_r+b_2\y\phi_r-\x\zeta_r-a_1b_2\x^2\psi_r)=0,
\end{eqnarray}
\begin{eqnarray}\label{4.162}
&&b_{21}'\x+b_{22}'\y+b_{23}'\z+g'(\xi_r+2a_1b_2\x\phi_r)+g(a_1'\y\zeta_r+2(a_1b_2)'\x\phi_r+2a_1'a_1b_2\x\psi_r)\nonumber\\
&+&(b_{21}+2a_1b_2g\phi_r)(a_{11}\x+(a_{12}+\alpha')\y+(a_{13}+\beta'\sin\alpha)\z\nonumber\\
&+&a_1f(\theta_r+a_2\y\phi_r-\x\zeta_r-a_1a_2\x^2\psi_r))
+(-\alpha'\x+\z\beta'\cos\alpha)(b_{22}+a_1g(\zeta_r+2a_1b_2\x\psi_r))\nonumber\\
&+&b_{23}((a_{31}-\beta'\sin\alpha)\x+(a_{32}-\beta'\cos\alpha)\y+a_{33}\z)\nonumber\\
&+&(\alpha'-a_{21}-2a_1a_2f\phi_r)(b_{11}\x+b_{12}\y+b_{13}\z+a_1g(\theta_r+b_2\y\phi_r-\x\zeta_r-a_1b_2\x^2\psi_r))\nonumber\\
&-&(a_{22}+a_1f(\zeta_r+2a_1a_2\x\psi_r))(b_{21}\x+b_{22}\y+b_{23}\z+g(\xi_r+2a_1b_2\x\phi_r))\nonumber\\
&-&(a_{23}+\beta'\cos\alpha)(b_{31}\x+b_{32}\y+b_{33}\z)\nonumber\\
&+&(b_{22}+a_1g(\zeta_r+2a_1b_2\x\psi_r))(a_{21}\x+a_{22}\y+a_{23}\z+f(\xi_r+2a_1b_2\x\phi_r))\nonumber\\
&-&(-1)^r\eta a_1^2g(\xi_r+2a_1b_2\x\phi_r)=0
\end{eqnarray} and
\begin{eqnarray}\label{4.163}
&&b_{31}'\x+b_{32}'\y+b_{33}'\z+b_{31}(a_{11}\x+(a_{12}+\alpha')\y+(a_{23}+\beta'\sin\alpha)\z\nonumber\\
&+&a_1f(\theta_r+a_2\y\phi_r-\x\zeta_r-a_1a_2\x^2\psi_r))\nonumber\\
&+&b_{32}((a_{21}-\alpha')\x+a_{22}\y+(a_{23}+\beta'\cos\alpha)\z+f(\xi_r+2a_1b_2\x\phi_r))
\nonumber\\
&+&(\beta'\sin\alpha-a_{31})(b_{11}\x+b_{12}\y+b_{13}\z+a_1g(\theta_r+b_2\y\phi_r-\x\zeta_r-a_1b_2\x^2\psi_r))\nonumber\\
&+&(\beta'\cos\alpha-a_{32})(b_{21}\x+b_{22}\y+b_{23}\z+g(\xi_r+2a_1b_2\x\phi_r))-a_{33}(b_{31}\x+b_{32}\y+b_{33}\z)\nonumber\\
&+&b_{33}(a_{31}\x+a_{32}\y+a_{33}\z)-b_{33}(\x\beta'\sin\alpha+\y\beta'\cos\alpha)=0.
\end{eqnarray}

 From the coefficients of $\x^2\phi_r$ in
$\Phi_{1\z}=\Phi_{3\x}$, we get
\begin{equation}\label{4.164}
(a_{23}+\beta'\cos\alpha)a_2f+\mu_0b_{23}b_2g=0.
\end{equation}
While, from the coefficients of $\x^2\phi_r$ in $\Psi_1=0$, we have
\begin{equation}\label{4.165}
-(a_{23}+\beta'\cos\alpha)b_2g+a_2fb_{23}=0.
\end{equation}
Since the functions $a_2$, $b_2$, $f$ and $g$ are nonzero, we obtain
that
\begin{equation}\label{4.166}
\left\{\begin{aligned}
   a_{23}&=-\beta'\cos\alpha,\\
   b_{23}&=0.
\end{aligned}\right.
\end{equation}
The coefficients of $\phi_r$ in $\Phi_{1\z}=\Phi_{3\x}$ suggest that
\begin{equation}\label{4.167}
a_{32}=-\beta'\cos\alpha.
\end{equation}
The coefficients of $\zeta_r$ in $\Phi_{2\z}=\Phi_{3\y}$ say that
\begin{equation}\label{4.168}
b_{32}=0.
\end{equation}
Thus the coefficients of $\xi_r$ in $\Psi_3=0$ imply
\begin{equation}\label{4.169}
\beta'=0.
\end{equation}
Substituting \eqref{4.166}-\eqref{4.169} into
\eqref{4.155}-\eqref{4.163}, we can simplify the calculation. One
shows that
\begin{equation}\label{4.170}
\left\{\begin{aligned} a_{13}&=a_{31}=0,\\
b_{ij}&=0\;\;\;\;\;\;\;\;\;\;\;\;\textrm{for }i,\ j=1,\ 2,\ 3,
\end{aligned}\right.
\end{equation}
\begin{equation}\label{4.171}
\left\{\begin{aligned}
   a_{21}=-a_{12}=\alpha'&=k,\\
   a_{11}=a_{22}=a_{33}&=0,
\end{aligned}\right.
\end{equation} and
\begin{equation}\label{4.172}
\left\{\begin{aligned}
  f&=c_1e^{(-1)^r\nu a_1^2t},\\
  g&=c_2e^{(-1)^r\eta a_1^2t},\\
  b_2&=a_2=\textrm{constant},
\end{aligned}\right.
\end{equation}
where $k$, $c_1$ and $c_2$ are constants. Thus one easily shows that
\begin{eqnarray}\label{4.175}
 &&p=\rho\{-a_1^2e^{(-1)^r2\nu a_1^2t}[\frac{1}{2}(4b\delta_{0,r}+c\delta_{1,r})c\x^2+a_2(2\delta_{0,r}(ab+c)+\delta_{1,r}c\cos b)\x^3\nonumber\\
 &+&a_1^2a_2^2(4a\delta_{0,r}+\delta_{1,r})\frac{\x^4}{2}+a_2(2\delta_{0,r}(ab-c)+\delta_{1,r}c\sin b)\x\y+(\delta_{1,r}cc_1\sin(b-b_1)\nonumber\\
 &+&2\delta_{0,r}(bc_1-b_1c))\x+\frac{a_1a_2}{2}(2\delta_{0,r}(c_1-ab_1)-\delta_{1,r}c_1\sin b_1)\x^2]+\frac{k^2}{2}\x^2\nonumber\\
 &-&e^{(-1)^r\nu a_1^2t}(a_1a_2\phi-2k)(\x\xi+a_1a_2e^{(-1)^r\nu a_1^2t}\x^2\phi_r)-(-1)^re^{(-1)^r2\nu a_1^2t}[a_e^{(-1)^r2\nu a_1^2t}\nonumber\\
 &\times&a_2e^{(-1)^r\nu a_1^2t}(a_1a_2\y\phi_r\psi_r-\frac{a_2}{2}(a_1(4a\delta_{0,r}+\delta_{1,r})\y^2+\phi_r^2)+\theta_r\zeta_r\nonumber\\
 &-&(\delta_{0,r}2(ab_1+c_1)+\delta_{1,r}c_1)\y)+k(\varepsilon_r+a_2\y\psi_r-\frac{a_2}{a_1}\phi_r)]-\frac{1}{2}e^{(-1)^r2\nu a_1^2t}\xi_r^2
 +\frac{k^2}{2}\y^2\}\nonumber\\
 &+&\mu_0\rho\{a_1c_2e^{(-1)^r\eta a_1^2t}(-a_1\xi_r\varepsilon_r+a_2\xi_r\phi_r-a_1a_2\y\xi_r\psi_r)\x\nonumber\\
 &+&((-1)^ra_1\xi_r^2-2a_1^2a_2\phi_r\varepsilon_r+2a_1a_2^2\phi_r^2-2a_1^2a_2^2\y\phi_r\psi_r)\frac{\x^2}{2}\nonumber\\
 &+&(-1)^ra_1^2a_2\x^3\xi_r\phi_r+(-1)^ra_1^3a_2^2\frac{\x^4}{2}\phi_r^2-\frac{a_1}{2}c_2e^{(-1)^r2\eta a_1^2t}(\theta_r+a_2\y\phi_r)^2
 -2a_1^2a_2c_2e^{(-1)^r2\eta a_1^2t}\nonumber\\
 &\times&(\delta_{0,r}[\frac{1}{2a_1}(b_1e^{2a_1\y}-a_1c_1e^{-2a_1\y})-(c_1+ab_1)\y]
 +\delta_{1,r}[\frac{c_1}{2a_1}\sin b_1\sin^2(a_1\y)\nonumber\\&+&\frac{c_1}{b_1}\cos b_1(\frac{a_1\y}{2}+\frac{\sin(2a_1\y)}{4})])\nonumber\\
 &-&2a_2^2c_2e^{(-1)^r2\eta a_1^2t}(\delta_{0,r}[-a_1^2\y^2+\frac{a_1\y}{2}(e^{2a_1\y}-e^{-2a_1\y})-\frac{1}{4}(e^{2a_1\y}+e^{-2a_1\y})]\nonumber\\
 &-&\delta_{1,r}[\frac{5}{8}\sin(2a_1\y)-\frac{\sin^2(a_1\y)}{2}-\frac{a_1\y\sin^2(a_1\y)}{2}+\frac{a_1\y}{4}])\}
\end{eqnarray} modulo the transformation in \eqref{T5}.

\begin{thm}
Let $k$, $l$, $a_1$, $a_2$, $c_1$ and $c_2$ be any real constants.
Then we have the following solutions of the MHD equations
\eqref{M1}-\eqref{M4}:
\begin{equation}\label{4.173}
   \left\{\begin{aligned}
       u=&k\x\sin(kt+l)-k\y\cos(kt+l)-c_1e^{(-1)^r\nu a_1^2t}(\xi_r+2a_1a_2\x\phi_r)\sin(kt+l)\\&+a_1c_1e^{(-1)^r\nu
           a_1^2t}(\theta_r+a_2\y\phi_r-\x\zeta_r-a_1a_2\x^2\psi_r)\cos(kt+l),\\
       v=&k\x\cos(kt+l)-k\y\sin(kt+l)+a_1c_1 e^{(-1)^r\nu
            a_1^2t}(\theta_r+a_2\y\phi_r-\x\zeta_r-a_1a_2\x^2\psi_r)\\
       &\times\sin(kt+l)+c_1e^{(-1)^r\nu
           a_1^2t}(\xi_r+2a_1a_2\x\phi_r)\cos(kt+l)\\
       w=&0,\\
   \end{aligned}\right.
\end{equation}
\begin{equation}\label{4.174}
   \left\{\begin{aligned}
       H^1=&a_1c_2e^{(-1)^r\eta
       a_1^2t}(\theta_r+a_2\y\phi_r-\x\zeta_r-a_1a_2\x^2\psi_r)\cos(kt+l)\\
       &-c_2e^{(-1)^r\eta a_1^2t}(\xi_r+2a_1a_2\x\phi_r)\sin(kt+l),\\
       H^2=&a_1c_2 e^{(-1)^r\eta a_1^2t}
       (\theta_r+a_2\y\phi_r-\x\zeta_r-a_1a_2\x^2\psi_r)\sin(kt+l)\\&+c_2 e^{(-1)^r\eta a_1^2t}(\xi_r+2a_1a_2\x\phi_r)\cos(kt+l),\\
       H^3=&0\\
   \end{aligned}\right.
\end{equation} and
$p$ is given by \eqref{4.175}.
\end{thm}

 \vspace{0.5cm}

 \textbf{Acknowledgement:} I would like to thank
Professor Xiaoping Xu for his advice and suggesting this research
topic.

\end{document}